\documentclass[11pt]{article}

\usepackage{amsfonts}
\usepackage{amsmath}
\usepackage{amssymb}
\usepackage{amsthm}
\usepackage{enumerate}
\usepackage{paralist}
\usepackage{fullpage}
\usepackage{hyperref}
\usepackage{tikz}
\usepackage{times}
\usepackage[margin=1in]{geometry}
\usepackage{paralist}
\allowdisplaybreaks

\usetikzlibrary{calc}

\newtheorem{theorem}{Theorem}[section]
\newtheorem{lemma}[theorem]{Lemma}
\newtheorem{proposition}[theorem]{Proposition}
\newtheorem{corollary}[theorem]{Corollary}
\newtheorem{definition}{Definition}

\newtheorem{remark}{Remark}

\newtheoremstyle{restate}{}{}{\itshape}{}{\bfseries}{~(restated).}{.5em}{\thmnote{#3}}
\theoremstyle{restate}
\newtheorem*{restate}{}

\newcommand{\avg}{\mathop{\mathrm{avg}}}

\newcommand{\E}{\mathop{\mathbb{E}}}

\newcommand{\cD}{\mathcal{D}}
\newcommand{\cX}{\mathcal{X}}
\newcommand{\cY}{\mathcal{Y}}
\newcommand{\cP}{\mathcal{P}}
\newcommand{\cL}{\mathcal{L}}
\newcommand{\cR}{\mathcal{R}}
\newcommand{\cO}{\mathcal{O}}

\newcommand{\unif}{\mathcal{U}}

\newcommand{\rev}[2]{\mathrm{rev}_{#1}(#2)}
\newcommand{\del}{\mathtt{delete}}
\newcommand{\ins}{\mathtt{insert}}
\newcommand{\con}{\mathtt{query}}

\newcommand{\Patrascu}{P\v{a}tra\c{s}cu}

\usepackage{color}

\newcommand{\SharedProbes}[1][v]{\left\lvert P_A(#1) \cap P_B(#1) \right\rvert}
\newcommand{\AllLevels}{[\log n]}
\newcommand{\TopLevels}{[L]}

\title{Cell-Probe Lower Bounds from Online Communication Complexity}
\author{Josh Alman\footnote{MIT CSAIL and EECS, jalman@mit.edu. Supported by an NSF Graduate Research Fellowship, and by NSF CAREER awards 1651838 and 1552651. Work initiated while at Stanford University.} \and Joshua R. Wang\footnote{Stanford University, joshua.wang@cs.stanford.edu. Supported by NSF CCF-1524062 and a Stanford Graduate Fellowship.} \and Huacheng Yu\footnote{Harvard University, yuhch123@gmail.com. Supported in part by NSF CCF-1212372.}}

\def\mainfile{}
\begin{document}
\begin{titlepage}
\maketitle

\begin{abstract}
  In this work, we introduce an online model for communication complexity.
  Analogous to how online algorithms receive their input piece-by-piece, our
  model presents one of the players, Bob, his input piece-by-piece, and has the
  players Alice and Bob cooperate to compute a result each time before the next piece is revealed to Bob.
 	This model has a closer and more natural correspondence to dynamic
  data structures than classic communication models do, and hence presents a
  new perspective on data structures.
  
  We first present a tight lower bound for the \emph{online set intersection} problem
  in the online communication model, demonstrating a general approach for
  proving online communication lower bounds. The online communication model
  prevents a batching trick that classic communication complexity allows, and
  yields a stronger lower bound. We then apply the online communication model to prove 
  data structure lower bounds for two dynamic data structure problems: the Group Range problem and the Dynamic Connectivity problem for forests.
  Both of the problems admit a worst case $O(\log n)$-time
  data structure. Using online communication complexity, we prove a tight
  cell-probe lower bound for each: spending $o(\log n)$ (even amortized) time per
  operation results in at best an $\exp(-\delta^2 n)$ probability of correctly
  answering a $(1/2+\delta)$-fraction of the $n$ queries.
\end{abstract}


\thispagestyle{empty}
\end{titlepage}

\clearpage

\ifx\undefined\mainfile



\fi

\section{Introduction}

One major hallmark of complexity theory is Yao's cell-probe model~\cite{Yao81},
a powerful model of computation that manages to capture the inherent complexity
found in a variety of data structure problems. The titular feature of this model
is that the data structure is only charged for \emph{the number of memory cells
that it accesses (or probes)}, and not for any computation it needs to perform on the contents
of those cells. Since this model is so strong -- data structures are given the power of `free
computation' --  proving lower bounds here yields lower bounds for
most other models of data structure computation. 
Many lower bounds in the cell-probe model are derived via connections
to communication complexity, wherein two players try to jointly compute a
function but are only charged for the bits that they communicate to each other
and again, not for any
computation~\cite{Ajtai88,MNSW95,BF02,Patrascu07,PT11,Yu16,WY16}.

Unfortunately, the sheer power granted to the data structure by the cell-probe model can
often make it difficult to prove strong lower bounds. In fact, in many cases a
matching lower bound in the cell-probe model may be impossible; counting just
cell probes in lieu of actual computation time might indeed make several
problems easier~\cite{LW17}. Partly as a result of this difficulty, only a few
techniques are known for proving cell-probe lower bounds. In this paper, we
propose a new technique to add to our growing toolbox. We propose a new model of
communication complexity which we call \emph{online communication}. We give tools
for proving lower bounds in this new model, and then use these tools to show how
the model results in new robust lower bounds for two fundamental data structure problems.

\subsection{Online Communication Model}

Inspired by the fact that a data structure must answer one query before it sees
the next, we propose a novel model of communication: the online communication
model. The salient feature of our model is that one of the players, Bob, does
not receive his entire input at once. Whereas Alice receives her entire input
$X$, Bob receives a small piece $Y_1$ and the two must jointly compute a
function $f_1(X, Y_1)$ before Bob receives the next piece $Y_2$, and so on. As
usual, we care about the total amount of communication that Alice and Bob use.
Intuitively, this model is designed to rule out batching techniques; in usual offline models of communication, it may be
cheaper for Bob to discuss all the pieces of his input together, but in the online model, this is impossible since he only receives one
piece at a time.

It stands to reason that we should be able to prove better lower bounds now that
communication protocols have one fewer trick to work with. In this paper, we develop
techniques that relate this model to more familiar entire-input-at-once models.
In order to demonstrate how things are different in the online model, we turn
to perhaps one of the most important communication problems: set disjointness.

\subsection{Online Set Disjointness}

The testbed for our new model is the quintessential problem, set disjointness. In the
basic version of this problem, Alice and Bob are each given subsets over $[n]$ and
want to compute whether their subsets are disjoint. This problem has long been
a favored source of hardness, and along with its many variations, it has been
thoroughly studied by theorists; see e.g. the surveys~\cite{chattopadhyay2010story, sherstov2014communication}.

In the context of our model, this problem manifests as the \emph{online set
intersection problem}. Alice is given an entire subset $X \subseteq [n]$ of
size $k$ while Bob is only given \emph{single} elements $y_i$ of another subset
$Y \subseteq [n]$ of size $k$ one at a time. The players need to decide whether
$y_i \not \in X$ before Bob receives the next element. We show that:
\begin{theorem}[informal]
  When $n \ge k^2$, the online set intersection problem requires
  $\Omega(k \log \log k)$ bits of total communication.
\end{theorem}
In fact, our proof shows that deciding whether $X$ and $Y$ are
disjoint requires $\Omega(k \log \log k)$ bits of communication in the online
model; it cannot be done more effiicently even if Alice and Bob may stop after finding an
intersecting element. We also give a fairly straightforward protocol which
solves the problem in $O(k \log \log k)$ bits of communication, showing that this bound is tight.
This stands in contrast to known bounds for the classical communication model, in which
the set disjointness problem can be solved
with just $O(k)$ bits by using a batching trick to test all elements of $Y$
simultaneously~\cite{HW07}.

\subsection{Group Range Problem}

The first data structure problem we consider is a generalization of the Partial
Sums problem (from e.g. \cite{puaatracscu2004tight}). In the \emph{Group Range
Problem}, we have a group $G$ along with a \emph{binary encoding} of the group
elements (any injective function) $e : G \to \{0, 1\}^s$. We would like a data
structure which stores a sequence of $n$ group elements $a_0, \ldots, a_{n-1}$
while supporting the following operations:
\begin{itemize}
  \item Update($i, a$) sets entry $a_i \leftarrow a$.
  \item Query($\ell, r, i$) returns the $i^{th}$ bit of the binary encoding of
    the group product $a_\ell a_{\ell + 1} \cdots a_{r - 1} a_r$.
\end{itemize}
We focus on the case where the cell-size is $w = \Theta(\log n)$
and the group is polynomially-sized: $\log |G| = O(w)$.

Regarding upper-bounds, there is a folklore data structure which solves the
problem with $O(\log n)$ time per operation. This is a worst-case (not just
amortized) guarantee, and the data structure is deterministic. There is a
matching $\Omega(\log n)$ cell-probe lower-bound by \Patrascu{} and Demaine for
the Partial Sums problem, wherein queries need to return the entire product
rather than a single bit~\cite{puaatracscu2004tight}. This lower bound holds for
Las Vegas randomized data structures (the number of cell probes is considered in
expectation) and amortized operation cost.

However, this lower bound leaves open several plausible ways to improve the
running time. What if we really only care about a single bit of each query? What
if we are willing to tolerate errors? Our main result shows that even if we
permit these concessions, the $\Omega(\log n)$ barrier still stands:
\begin{theorem}\label{thm:group_range_LB}
There exists a distribution over $n$ updates and queries for the Group Range
Problem with binary encoding of the group elements $e:G\rightarrow \{0,1\}^s$,
such that for any randomized cell-probe data structure $D$ with word size
$w = \Theta(\log n)$, which with probability $p$ answers at least a
$(\frac{1}{2} + \delta)$ fraction of queries correctly and spends $\epsilon n \log n$
total running time, we must have $p \leq \exp(-\delta^2 n)$, as long as
$s \leq (1+\epsilon)\log |G|$, $\delta^2\gg\epsilon\geq \Omega(1/\log n)$, and
$n$ is sufficiently large.\footnote{In this paper, we use $\exp(f(n))$ to mean $2^{\Theta(f(n))}$.}
\end{theorem}

Put another way, Theorem~\ref{thm:group_range_LB} settles the trade-off between
running time and accuracy of the output for the Group Range Problem. There are
two possible regimes. If we are willing to pay $\Theta(\log n)$ time per
operation, then there exists a deterministic worst-case data structure.
Otherwise, if we require the data structure to spend $o(\log n)$ time per
operation, then Theorem~\ref{thm:group_range_LB} shows that we cannot
hope to do much better than outputting a random bit for each query, up to a
constant factor improvement in $\delta$. To the best of our knowledge, this
bound and our other lower bound we describe shortly are the first tight data
structure lower bounds in such a high error regime, where a data structure may
answer barely more than half of the queries correctly, and do so even with a
small success probability.

\subsection{Dynamic Connectivity}

Next, we consider a fundamental problem in graph data structures: Dynamic
Connectivity. In this problem, we would like a data structure which stores an
undirected graph $G = (V, E)$ on $n$ vertices, while supporting the following
operations:
\begin{itemize}
  \item $\ins(u, v)$ adds edge $(u, v)$ to $E$.
  \item $\del(u, v)$ removes edge $(u, v)$ from $E$.
  \item $\con(u, v)$ returns whether or not there currently exists a path between
    nodes $u$ and $v$.
\end{itemize}

Like before, we look at this problem in the cell-probe model with cell size $w = O(\log n)$.
The link/cut tree data structure~\cite{sleator1981} and Euler tour tree data structure~\cite{henzinger1999randomized} for the problem take $O(\log n)$ time per update or query.
A matching $\Omega(\log n)$ lower bound was given by \Patrascu{} and Demaine~\cite{PD06}.
However, their lower bound holds for Las Vegas or Monte Carlo data structures with amortization, where they assume that the error probability for each query is $n^{-c}$ for some large constant $c$.

\Patrascu{} and Demaine still leave open the question of what can be done if we insist on $o(\log n)$ time per operation.
Their lower bound asserts that we cannot answer each query correctly with better than $1-n^{-c}$ probability.
However, for one example, it could still be possible to design a data structure which answers \emph{all} queries correctly simultaneously with probability, say $1-1/\log n$, and such that each individual query is correct with probability lower than $1-n^{-c}$.
Such a data structure would not violate the existing lower bounds, and its success probability would be good enough in many applications, as it only incurs an additive $1/\log n$ overall error probability.

Again, our new technique yields a robust lower bound, ruling out such data structures:
\begin{theorem}\label{thm:dynamic_graph_LB}
There exists a distribution over $O(n)$ updates and queries for the dynamic connectivity problem, such that for any randomized cell-probe data structure $D$ with word-size $w=\Theta(\log n)$, which with probability $p$ answers at least a $(\frac{1}{2}+\delta)$-fraction of the queries correctly and spends $\epsilon n\log n$ total running time, we must have $p\leq \exp(-\delta^2n)$ as long as $\delta^2\gg 1/\log(1/\epsilon)$ and $\epsilon\geq \Omega(1/\log n)$ and $n$ is sufficiently large.
 Moreover, the graph is always a forest throughout the sequence of updates.
\end{theorem}

Similar to before, this essentially settles the complexity of the Dynamic Connectivity problem in forests (where the graph is always a forest throughout the sequence of updates). If one wants $o(\log n)$ per update and query, then one cannot do better than outputting the flip of a random coin to answer each query, again up to a constant factor in $\delta$.

Our lower bound almost matches the best known upper bound for Dynamic Connectivity in general graphs of $O(\log n \log^3 \log n)$ by Thorup \cite{thorup2000}. 
Dynamic Connectivity with higher error than \Patrascu{} and Demaine allowed for, although still lower error than we consider, was studied by Fredman and Saks \cite{FS89}, but for worst-case update time instead of amortized, and for the problem where edge deletions are not allowed; the only updates allowed are edge insertions. 
They showed that any $1/32$-error data structure for Dynamic Connectivity without deletions with expected query time $t_q$ and \emph{worst-case} update time $t_u$ must have $t_u \geq \Omega(\log(n) / \log(t_u \log (n)))$. Ramamoorthy and Rao~\cite{RR16} recently gave a simplified proof of Fredman and Saks' result as well.

\subsection{Further Results}

We also prove some complementary results to our two data structure lower bounds.

\subsubsection{Group Range Problem}

The Group Range Problem is stated very broadly about general groups $G$.
Although it may help the reader to imagine a more common group like
$\mathbb{Z}_m$ or a permutation group while reading the proof, there are other
important cases. For example, Theorem~\ref{thm:group_range_LB} holds when
$G$ is the direct product of many smaller groups. In this case, the problem can
be viewed as many disjoint copies of the Group Range problem on the smaller
component groups with simultaneous updates.

The case where $G$ is the general linear group of invertible matrices also has
many applications; see Appendix \ref{app:app} for a discussion of applications
to physics and to other dynamic data structure problems. For this case, we show
how the matrix structure can be exploited to prove even stronger results. For
example, as a variant of the original problem, consider the \emph{Matrix Product
Problem}, where queries can only ask for a bit about the bottom-right entry of
the product of the \emph{entire} range of matrices, rather than any bit about
the product of any subrange. In Appendix~\ref{s:matrix} we show that the lower
bound still applies:
\begin{corollary}\label{cor:matrix-prod-lower}
  Theorem~\ref{thm:group_range_LB} holds for the Matrix Product Problem.
\end{corollary}
We show a similar result for upper-triangular matrices as well in
Appendix~\ref{subsec:upper-tri}.

It would be interesting to extend Theorem~\ref{thm:group_range_LB} to hold for
an even wider class of algebraic structures. For instance, some past work
(e.g.~\cite{puaatracscu2004tight}) considers the partial sums problem where $G$
is any \emph{semi-group}. However, we show that such an extension is impossible, not only
to semi-groups, but even just to monoids (a type of algebraic structure between
groups and semi-groups, which satisfies all the group axioms except the
existence of inverses). Indeed, we demonstrate in Appendix~\ref{ss:monoid} that the
$\Omega(\log n)$ lower bound can be beaten for the \emph{Monoid Range Problem}
(the same as the Group Range Problem except that $G$ can be any monoid), so no
general lower bound applies:
\begin{theorem}\label{thm:monoids}
  There exists a family of monoids $(G_n)_n$ such that the Monoid Range Problem
  can be solved in $O \left( \frac{\log n}{\log \log n} \right)$ time per operation
	worst-case deterministically in the cell-probe model.
\end{theorem}

\subsubsection{Dynamic Connectivity}

Dynamic Graph Connectivity is one of the most basic and versatile dynamic graph problems. As such, we can extend Theorem~\ref{thm:dynamic_graph_LB} to hold for a number of other graph problems. Some examples include:

\begin{itemize}
\item Dynamic Entire Graph Connectivity: Maintain a dynamic undirected graph, where queries ask whether the entire graph is connected.
\item Dynamic Minimum Spanning Forest:  Maintain a dynamic undirected graph, where queries ask for the size of a minimum spanning forest.
\item Dynamic Planarity Testing: Maintain a dynamic undirected graph, such that edge insertions are guaranteed to maintain that the graph is planar, and where queries ask whether inserting a specific new edge would result in a non-planar graph.
\end{itemize}

\begin{corollary} \label{cor:connectivity}
Theorem~\ref{thm:dynamic_graph_LB} holds for Dynamic Entire Graph Connectivity, Dynamic Minimum Spanning Forest, and Dynamic Planarity Testing.
\end{corollary}

Corollary \ref{cor:connectivity} follows from some straightforward reductions given in \cite[Section 9]{PD06}.

\subsection{Our Technique and Related Work}
Next, we discuss our plan of attack for using the online communication model along with
other ideas to prove our data structure lower bounds, 
Theorems~\ref{thm:group_range_LB}~and~\ref{thm:dynamic_graph_LB}, and we compare it with
the approaches of past work.
A more detailed overview of our proofs is given later in Section \ref{sec:overview}.

Our high-level
strategy is similar to previous techniques based on communication complexity for proving
cell-probe lower bounds~\cite{PT11,Yu16,WY16}. We first ``decompose'' the
computation being done into several communication games, and show that an
efficient data structure would induce efficient protocols for these games. We
then prove communication lower bounds for these games, ruling out these supposed
efficient protocols. The communication games we wind up with consist of
a random sequence of interleaved updates and queries divided into two
consecutive blocks of operations. Roughly speaking, in each
communication game, the first block is only revealed to Alice while the second
block is only revealed to Bob. All other operations are revealed to both
players. The goal of the game is for Alice and Bob to cooperatively answer all 
the queries in Bob's interval.

The choice of what communication model to use in this strategy is crucial.
The first step, transforming a fast data structure into an efficient protocol,
can be done more efficiently in a stronger model (e.g. randomized over deterministic). On the other hand, the second
step, proving communication lower bounds, is more difficult in a stronger model.
Designing the right communication model to balance these two proof phases is a crucial
ingredients in these types of proofs.

In this paper, we analyze the communication games in our online communication model.
Compared to other models used in previous work,
our model has a more natural correspondence to the task that data structures
face: answering queries in sequential order. Studying these communication games
in our online communication model yields a more fine-grained view of
the situation. See Section~\ref{sec:overview} and Section~\ref{sec_step3} for more details on this connection
between online communication complexity and dynamic data structures.

\subsubsection{Group Range Problem}

To illustrate this point, consider the communication games induced by the Group
Range Problem. When one analyzes these games in the classical communication models considered by past work, where
both players receive their inputs at once, there is a protocol which is too
efficient to prove a tight lower bound\footnote{For some $G$, Bob has a succinct encoding of his queries and
can send the compressed input to Alice in order to solve the problem more
efficiently than the trivial protocol would.}. In other words, it is
provably impossible to use any of the previous communication models at this
point in the proof; the communication game is simply not ``hard'' in any of
them. We will see that these communication games are hard enough to prove strong lower bounds in our online communication model.

As stated before, \Patrascu{} and Demaine~\cite{puaatracscu2004tight} proved a
$\Omega(\log n)$ lower bound for the Partial Sums problem (queries want entire
product rather than a single bit), when no error is allowed in answering
queries. Their \emph{information-transfer technique} does not apply directly to
our problem, since it relies on the fact that each query outputs many bits and
hence reveals a lot of information, and that the data structure has no errors.
Their technique was later generalized~\cite{PD06} to prove lower bounds for
problems with single-bit output, but their argument mostly focuses on the query
which the data structure spends the least amount of time on. It is hard to apply
this generalization directly when both overall running time and overall accuracy
need to be taken into account. However, it is worth noting that their argument
does apply to our Group Range Problem if only zero-error data structures are
considered.

\subsubsection{Dynamic Connectivity}

The high-level structure of our Dynamic Connectivity lower bound proof is close to that of P\v{a}tra\c{s}cu and Demaine's proof~\cite{PD06}.
To prove an $\Omega(n\log n)$ lower bound on the total running time on $O(n)$ operations, both proofs reduce the task to proving that given an initial graph, $k$ updates and $k$ queries, if we perform the updates on the initial graph and then ask the $k$ queries, 
then there must be a big, $\Omega(k)$-size intersection between the set of cells probed and written to during the insertions, and the set of cells probed during the queries.
Intuitively, we need to show that the data structure must learn enough information about the updates in order to answer the queries.

The two proofs then diverge from this point onwards.
P\v{a}tra\c{s}cu and Demaine first set up a hard distribution on updates and queries such that when the answers to all $k$ queries are \texttt{Yes}, one is able to reconstruct the $k$ updates exactly based on these queries.
Then they use an encoding argument to show that if the data structure only probes $o(k)$ such cells, then the $k$ updates can be encoded very efficiently, contradicting an information theoretical lower bound that they prove using the distribution itself.
Roughly speaking, they encode the $k$ updates so that one is able to ``simulate'' a data structure on any sequence of $k$ queries after the updates based only on the encoding. 
Then one can iterate over all possible queries, simulate the data structure on all of them, and find the one with $k$ \texttt{Yes} answers, which can be used to reconstruct the updates.

Since the information about updates is only hidden in the all-\texttt{Yes} queries, and one needs to simulate on a large number of queries before $k$ \texttt{Yes} queries are found, P\v{a}tra\c{s}cu and Demaine's argument fails if the data structure is allowed high two-sided error.
In fact, their proof only applies to the case where the error probability of each query is $1/n^c$ for some large constant $c$. 
It is not hard to prove that under their input distribution, one will not be able to learn much from the simulations if the error probability of each query is higher than about $1/\sqrt{n}$.

In order to resolve this issue in our Dynamic Connectivity lower bound, we first construct a different hard distribution such that not only the all-\texttt{Yes} queries, but even a random set of queries reveals a sufficient amount of information about the updates with high probability.
To prove our lower bound, we then use a very different encoding argument, based on the transcript of an online communication protocol.
We prove that if an efficient data structure exists, then there is an efficient online communication protocol for the problem where Alice receives the $k$ updates, Bob receives the $k$ queries one at a time, and the goal is to answer all queries.
Our encoding argument is more similar to those used in~\cite{CGL15} and~\cite{WY16}.
See Section~\ref{subsec:dslb-overview} for a more detailed overview of our approach.

Fredman and Saks~\cite{FS89} and Ramamoorthy and Rao~\cite{RR16} proved a lower bound for the insert-only version of Dynamic Connectivity, where deletion updates are not allowed.
They proved that for data structures with worst-case update time and constant error probability, the insert-only version of the problem has to take $\Omega(\log n/\log\log n)$ time per operation.
However, the insert-only regime is very different from our fully dynamic regime.
For worst-case data structures, the $\log n/\log\log n$ bound is tight~\cite{Blum85,Smid90}.
If we allow amortization, the standard union-find solution solves the problem in $O(\alpha(n))$ time per operation.
Thus, it is difficult to apply their technique to the fully dynamic regime.

\subsection{Organization}

We first formally define the online communication model in
Section~\ref{sec:model}, and then in Section~\ref{sec:overview} we give an overview of all three of our lower bound proofs.
Thereafter we prove our main results:
we prove the online set intersection lower
bound in Section~\ref{sec_disj}, then in Section~\ref{s:group-range-section} we prove the cell-probe lower bound for the Group Range Problem, and in Section~\ref{s:dynamic-connectivity}
we prove the cell-probe lower bound for Dynamic Connectivity. Finally,
in Appendix~\ref{s:further-group-results} we prove our further results about the Group Range Problem.

\ifx\undefined\mainfile
\bibliography{matrix-seq}
\bibliographystyle{alpha}
\end{document}
\fi

\section{The Online Communication Model}
\label{sec:model}

In this section, we define the online communication model, and then throughout the rest of the paper we present some approaches for proving lower bounds in it. We intentionally try to keep the model quite general. In Section~\ref{sec_disj}, we showcase our approach by proving a \emph{tight} lower bound for the natural variation of set-intersection for this setting, and thereafter we use the model to prove cell-probe data structure lower bounds.

In the online communication model, there are two players, Alice and Bob. Alice
is given her entire input $X \in \cX$ at the beginning. Bob will be given his
input $Y_1, Y_2, \ldots, Y_k \in \cY$ gradually. The two of them want to compute
$f_1(X, Y_1), f_2(X, Y_2), \ldots, f_k(X, Y_k)$ under the
following circumstances:
\begin{enumerate}
	\item
		The game consists of $k$ stages.
		The players remember the transcript from previous stages.
	\item
		At the beginning of Stage $i$ for $i \in [k]$, $Y_i$ is revealed to \emph{Bob}.
	\item
		Next, the players communicate as if they were in the classical communication model.
		After that, Bob must output $f_i(X, Y_i)$.
	\item
		At the end of Stage $i$, $Y_i$ is revealed to \emph{Alice}, and the players proceed to the next stage.
\end{enumerate}

Note that the number of stages, $k$, is fixed and known up-front when designing a protocol. In a deterministic (resp. randomized) online communication protocol, the players communicate as if they were in the deterministic (resp. randomized) communication model in the second step of each stage.

We desire protocols that use the minimum amount of \emph{total} communication. A protocol is free to perform a different amount of communication in each stage. However, there is a natural tension on the proper time to communicate: in earlier stages the players have less information, but they still need to solve their current task at hand before they can proceed. Later on, we will see that the total communication will correspond nicely with the amortized cost of data structure operations.

\ifx\undefined\mainfile


\fi

\section{Proof Overviews}
\label{sec:overview}

\subsection{Online Set Intersection Lower Bound}
In this section we give a high-level overview of how we prove our communication lower bound for online set intersection (OSI). Although the lower bound for OSI is not explicitly used in our data structure lower bounds later, the data structure lower bounds do use online communication lower bounds for other problems which we prove using some common techniques. Our OSI lower bound is, in a sense, a warm-up for the more complex proofs to come.

The main idea behind our OSI lower bound is a very general reduction showing how online communication lower bounds can be proved using techniques from offline communication lower bounds. Consider an offline communication problem called the Index problem, where Alice is given a set $X \subseteq \{1,2,\ldots,n\}$ of size $|X|=k$, and Bob is only given a single element $y \in \{1,2,\ldots,n\}$, and their task is to determine whether $y \in X$. One can view the OSI problem as $k$ iterations of Alice and Bob solving the offline Index problem.

That said, it is insufficient to simply prove a lower bound for the Index problem. Since Alice has the same set $X$ in all $k$ iterations, Bob can learn information about it throughout the rounds of the protocol, and so it is plausible that later rounds can be completed with less communication than earlier rounds. In order to circumvent this issue, we prove:

\begin{lemma} \label{lem:index-informal}
(informal) There is a protocol for OSI which in total uses $g(n,k)$ bits of communication in expectation \emph{if and only if} there is a protocol for Index where
\begin{compactenum}
\item Alice first sends $O(g(n,k))$ bits in expectation, then
\item Alice and Bob speak an additional $O(g(n,k) / k)$ bits in expectation.
\end{compactenum}
\end{lemma}
The high-level idea for proving the `only if' direction of Lemma~\ref{lem:index-informal} is that Alice can begin by telling Bob a cleverly-crafted message containing the information that Bob would learn about $X$ during the OSI protocol on a random sequence of inputs. Thereafter, Alice and Bob can pretend they are in the `easiest' round of their OSI protocol, which only takes $O(g(n,k) / k)$ bits in expectation to solve. Once we prove Lemma~\ref{lem:index-informal}, it remains to prove a lower bound for the Index problem in the usual offline communication model, which can be doine using standard counting techniques.

We actually prove a more general version of Lemma~\ref{lem:index-informal} for any online communication problem in which Alice and Bob are computing the same function $f = f_i$ in each round (in the case of OSI, $f$ is the Index problem). Unfortunately, for our data structure lower bound proofs, the communication games do not have this property, and more care will be needed.

\subsection{Data Structure Lower Bounds} \label{subsec:dslb-overview}
In this section, we give a streamlined overview of our data structure lower bound proofs.
The proofs of our lower bounds for Group Range and Dynamic Connectivity both have a similar high-level structure.
In both proofs, the first step is to design a hard input distribution.
The distribution is supported on sequences of operations consisting of $O(n)$ mixed updates and queries.
Then by Yao's minimax principle~\cite{yao1977probabilistic}, it suffices to prove a lower bound against \emph{deterministic} data structures dealing with inputs drawn from this distribution.

Next, to prove a lower bound of $\Omega(n\log n)$ for answering the random sequence, we use an idea from~\cite{PD04}, which reduces proving a lower bound on total running time to proving lower bounds for many subproblems.
Each subproblem is defined by two adjacent intervals of operations of equal length from this random sequence, which are denoted by $I_A$ and $I_B$, e.g., $I_A$ is the interval consisting of the 17th to the 32nd operation in the sequence, and $I_B$ is the interval consisting of the 33rd to the 48th operation.
In each subproblem, instead of the running time (i.e., the number of cell-probes), we are interested in the number of cells that are probed in \emph{both} intervals $I_A$ and $I_B$.
A counting argument from~\cite{PD04} shows that
\vspace{-7pt}
\begin{itemize}
\item \emph{if} for every $k$ and adjacent interval pair $(I_A, I_B)$ of length $k$, at least $\Omega(k)$ cells are probed in both $I_A$ and $I_B$, \vspace{-7pt}
\item \emph{then} the total running time is at least $\Omega(n\log n)$.
\end{itemize}
\vspace{-7pt}
In order to prove a lower bound when the data structure's goal is only to maximize the probability of answering $(1/2+\delta)$-fraction of the queries correctly, we generalize the argument, and show that
\vspace{-7pt}
\begin{itemize}
\item \emph{if} for every $\delta',k$ and adjacent interval pair $(I_A, I_B)$ of length $k$, the probability that $o(k)$ cells are probed in both $I_A$ and $I_B$ and $(1/2+\delta')$-fraction of the queries in $I_B$ is correct is $\exp(-\delta'^2 k)$, \vspace{-7pt}
\item \emph{then} the probability that total running time is $o(n\log n)$ and $(1/2+\delta)$-fraction of the queries is correct in total is $\exp(-\delta^2 n)$.
\end{itemize}
\vspace{-7pt}
Thus, the tasks boil down to proving such lower bounds for all subproblems.

\subsubsection{Online Communication Simulation}
We now focus on a single subproblem $(I_A, I_B)$.
We would like to show that if a data structure answers a $(1/2+\delta)$-fraction of the queries in $I_B$ correctly, then it must probe many cells in $I_B$ which were also probed and written to in $I_A$.
Intuitively, if a data structure probes very few cells in $I_B$ that are probed in $I_A$, then it learns very little information about the updates in $I_A$.
Thus, if the answer to a random query would reveal one bit of information about the updates in $I_A$, but the data structure has learned a negligible amount of information about $I_A$, then the data structure cannot hope to answer the query with a nonnegligible advantage above $1/2$.
To formulate the above intuition, we model this process by an online communication game.

\paragraph*{Communication Game.}
We define one communication game for each interval pair $(I_A, I_B)$.
Fix two intervals $I_A = I_A(v)$ and $I_B = I_B(v)$ consisting of $k$ updates and
queries each, all the operations $O$ prior to these intervals, all the queries $Q_A$ in
$I_A$ and all the updates $U_B$ in $I_B$. That is, the only undetermined operations
up to the end of $I_B$ are the updates in $I_A$ and the queries in $I_B$; everything else is common knowledge to Alice and Bob. We
embed these undetermined operations into a communication game. In the associated online
communication game $G = G(v, O, Q_A, U_B)$, $X$ consists of the updates in $I_A$,
and $Y_i$ is the $i^{th}$ query in $I_B$. The goal of Stage $i$ is to compute
the $i^{th}$ query in $I_B$.

Now we present (an informal version of) our main lemma, which connects the data structures to online communication.

\begin{lemma}[informal]
For any data structure $D$, there is a protocol $\cP_D$ for the communication game
$G(v, O, Q_A, U_B)$ such that
\begin{enumerate}
	\item
		Bob sends no message;
	\item
		For every $\beta\in(0,1)$, the probability that 
		\vspace{-3pt}
		\begin{itemize}
			\item Alice sends $o(k\log n)$ bits, and
			\item $\cP_D$ answers a $(\beta-o(1))$-fraction of the $f_i(X,Y_i)$'s correctly
		\end{itemize}
		is at least the probability \emph{conditioned on} $O,Q_A,U_B$ that
		\vspace{-3pt}
		\begin{itemize}
			\item $o(k)$ cells are probed in both $I_A$ and $I_B$ by $D$, and
			\item $D$ answers a $\beta$-fraction of queries in $I_B$ correctly.
		\end{itemize}
\end{enumerate}
\end{lemma}

For any data structure $D$, we construct the protocol $P_D$ as follows.
\begin{enumerate}
	\item (Preprocessing)
		Recall that Alice knows all the operations up to the end of $I_A$ and the updates in $I_B$,
		and Bob knows all the operations prior to $I_A$ and all the operations in $I_B$.
		First, Alice simulates $D$ up to the end of $I_A$, and Bob simulates $D$ up to the beginning of $I_A$ and \emph{skips $I_A$}.
		Denote the memory state that Alice has \emph{at this moment} by $M_A$.
		Next, the players are going to simulate operations in $I_B$.
	\item (Stage $i$ - Alice's simulation)
		Since the $(i-1)$-th query is revealed to Alice at the end of the last stage, Alice first continues her simulation of $D$ up to right before the $i$-th query.
		Alice then sends Bob the cells (their addresses and contents in $M_A$) that are
		\begin{itemize}
			\item
				probed during this part of the simulation, and
			\item
				probed during $I_A$, and
			\item
				not probed in the previous stages.
		\end{itemize}
	\item (Stage $i$ - Bob's simulation)
		Bob first updates his memory state according to Alice's message: For each cell in the message, Bob replaces its content with the actual content in $M_A$.
		This is the first time $D$ probes these cells, since otherwise Alice would have sent them earlier, and so their contents remain the same as in $M_A$. 
		Bob then continues his simulation of $D$ up to the beginning of $i$-th query.
	\item (Stage $i$ - query answering)
		Bob now simulates $D$ on query $Y_i$. 
		During the simulation, Bob \emph{pretends} that he has the right memory state of $D$ for the query, even though he skipped $I_A$, and has only received partial information from Alice about it.
		He then outputs whatever answer $D$ gives him.
		Finally, Bob rolls back his copy of $D$ to the version right before this query (after the simulation described in the previous step). 
		Even though he was assuming his copy of $D$ is correct, it may have actually made a mistake, and at the beginning of the next step, Alice will tell Bob what cells he should have queried and changed.
\end{enumerate}

The key observation to make about the above protocol is that Bob might only give a different answer to query $Y_i$ than the real data structure $D$ would have if $D$ would probe a cell that was written to during $I_A$ while answering $Y_i$. Moreover, at the beginning of the next stage, Alice would then tell Bob the true value which that cell should have had. Hence, each cell which $D$ would write to in $I_A$ and probe in $I_B$ can cause Bob to make at most one mistake. As such, if $D$ would only probe a negligible number, $o(k)$ of cells in both $I_A$ and $I_B$, then Bob similarly gives the same answer as a correct $D$ would to all but a negligible number of his queries.

\subsubsection{Online Communication Lower Bounds}
The tasks now reduce to proving online communication lower bounds.
We prove the communication lower bounds for Group Range and Dynamic Connectivity using different approaches.

\paragraph{Communication lower bounds for Group Range.}
We design the hard distribution such that the $k$ updates in $I_A$ have entropy about $k\log n$.
Hence, if Alice sends only $o(k\log n)$ bits to Bob, then Bob knows very little about those updates.
In particular, we carefully design the queries such that there is a $\Theta(k\log n)$-bit encoding of the $k$ updates, and each query is essentially asking for one random bit of this encoding.
Then on average, every bit is still close to unbiased even after Bob sees Alice's message.
That is, Bob will not be able to predict the answer with much better probability than $1/2$.

Furthermore, we prove the above \emph{conditioned on} whether Bob answered the previous queries correctly.
Therefore, the sequence of numbers consisting of, for each $1\leq i\leq k$, the number of correct answers in the first $i$ queries \emph{minus} its expected value, forms a supermartingale.
Applying the Azuma-Hoeffding inequality shows that the probability that at least a $(1/2+\delta)$-fraction of the queries is correct is at most $\exp(-\delta^2 k)$.

\paragraph{Communication lower bounds for Dynamic Connectivity.}
The lower bound for the Dynamic Connectivity problem is proved in a different way.
To prove the communication lower bound, we first show that it suffices to prove that the probability that all $k$ queries are correct is at most $2^{-(1-o(1))k}$.
This would in particular imply that the probability that all $k$ queries are wrong is also at most $2^{-(1-o(1))k}$.
In fact, for any fixed sequence of choices of whether each query is correct or not, we show that this sequence happens with probability no more than $2^{-(1-o(1))k}$. This is, in particular, at most a $2^{o(k)}$ factor more than the probability of achieving the fixed sequence by outputting uniformly independent bits.
This implies that the probability that a $(1/2+\delta)$-fraction of the queries is correct is at most $2^{o(k)}$ times the probability of the same event when all the bits are independent, which is $\exp(-\delta^2 k)$.

Next, we prove that when the inputs to the communication problem are independent and Bob does not speak, we may assume without loss of generality that Alice only speaks before the first stage,\footnote{Note that even if Bob does not speak during an online communication protocol, Alice still learns what Bob's inputs are each time Bob finishes answering a query.} which we call the Monologue lemma:
\begin{lemma}[Monologue Lemma (informal)]
  Suppose that Alice's input $X$ and Bob's inputs $Y_1, \ldots, Y_k$ are independent, and there is a protocol $P$ such that:
  \begin{compactenum}
    \item Only Alice talks.
    \item At most $C$ bits are sent.
    \item All $k$ queries are answered correctly with probability $p$
  \end{compactenum}
  Then there is another protocol $P'$ with the following properties:
  \begin{compactenum}
    \item Only Alice talks, and she only does so in the first stage.
    \item $C + O(\log 1/p)$ bits are sent in
      expectation.
    \item All $k$ queries are answered correctly with probability at least $p$.
  \end{compactenum}
\end{lemma}

Using this lemma, we will be able to prove the communication lower bound.
Assume for the sake of contradiction that Alice sends $o(k\log n)$ bits and Bob answers $k$ queries correctly with probability at least $2^{-0.99k}$.
The high-level idea is to let Alice simulate the protocol and send a message about her input, which takes $o(k\log n)$ bits.
Since Bob is able to complete the protocol with no further communication, we know that a random sequence of $k$ queries can be answered correctly based solely on this message with probability $2^{-0.99k}$.
The players then treat the public random string as repeated samples of queries.
On average, there is one entirely-correct sample of queries in every $2^{-0.99 k}$ samples from the public randomness.
Thus, it only takes about $0.99 k$ bits for Alice to specify each sample that would be answered entirely correctly by Bob.
Ideally, each of these samples of $k$ queries reveals $k$ bits of information about Alice's input.
That is, in the ideal situation, Alice will be able to save about $0.01 k$ bits each time at the cost of sending $o(k\log n)$ extra bits in the beginning.
If Alice managed to repeat this much more than $0.001\log n$ times, and each time revealed about $k$ extra bits of information, she would have revealed $0.001k\log n$ bits of information in total using only $(0.00099+o(1))k\log n$ bits, which yields a contradiction.

\ifx\undefined\mainfile
\bibliography{matrix-seq}
\bibliographystyle{alpha}
\end{document}
\fi

\ifx\undefined\mainfile



\fi

\section{Online Set-Intersection Lower Bound}
\label{sec_disj}

\paragraph{Online Set-Intersection.}
In the online set-intersection problem (OSI), Alice is given one set $X$ of size $k$ over the universe $[n]$.
In each stage, Bob is given an input $Y_i\in [n]$, which is an element in the same universe.
The goal of this stage is to verify whether $Y_i\in X$. 
Equivalently, the inputs are two (multi-)sets $X, Y\subseteq [n]$ of size $k$ each. 
Each element of the set $Y$ is revealed one by one.
The goal is to compute their intersection. 

\begin{theorem}\label{thm_OSI_LB}
For $n\geq k^2$, any zero-error OSI protocol using public randomness must have expected total communication cost at least $\Omega(k\log\log k)$.
\end{theorem}

It is not hard to see that $\Omega(|X \cap Y| \log n)$ is also a lower bound, since Alice and Bob need to confirm that their elements in common are actually equal; in other words, our combined lower bound is $\Omega(k \log \log k + |X \cap Y| \log n)$. Before we prove Theorem~\ref{thm_OSI_LB}, we give a protocol which shows that this bound is tight.

\begin{lemma} \label{lem:almost-tight}
There is a zero-error OSI protocol using public randomness with expected communication cost $O(k \log \log k + |X \cap Y| \log n)$.
\end{lemma}
\begin{proof}
The protocol is as follows:
\begin{enumerate}
  \item
    The players use public randomness to sample two uniformly random hash functions $h_1: [n]\rightarrow [k^{2}]$ and $h_2 : [k^{2}] \to [k \log k]$, and define $h : [n] \to [k \log k]$ by $h = h_2 \circ h_1$.
  \item
    Alice sends Bob the set $h(X)$ in $O(\log {k\log k\choose k})=O(k\log\log k)$ bits\footnote{Recall that for any integers $n\geq m > 0$ we have $\binom{n}{m} \leq \left( \frac{n \cdot e}{m} \right)^m$. Hence, $\binom{k \log k}{k} \leq O(\log k)^k$.}.
  \item
    For each $Y_i$:
		\begin{enumerate}
		\item If $h(Y_i)$ is not in $h(X)$, Bob returns ``NO'' immediately.
		\item\label{commstagetwo} Otherwise, Bob sends Alice $h_1(Y_i)$, and Alice tells Bob whether it is in $h_1(X)$. If not, Bob returns ``NO'' immediately.
		\item\label{commstagethree} Otherwise, for each $X_j \in X$ such that $h_1(X_j) = h_1(Y_i)$, Alice and Bob determine whether $X_j = Y_i$. They do this with the zero-error protocol for equality which uses $O(\log n)$ bits of communication if $X_j = Y_i$ and $O(1)$ bits of communication in expectation if $X_j \neq Y_i$. If $X_j = Y_i$ they return ``YES'', and if $x_j \neq Y_i$ for each such $X_j \in X$, they return ``NO''.
		\end{enumerate}
\end{enumerate}
For each $Y_i\notin X$, the probability that $h(Y_i) \in h(X)$ is at most $1/\log k$. Since it takes $O(\log k)$ bits for Bob to send $h_1(Y_i)$ to Alice, the total expected communication cost for stage \ref{commstagetwo} over all $i$ with $Y_i \notin X$ is $O(k)$. Similarly, for each $Y_i \notin X$, the expected number of $X_j \in X$ such that $h_1(X_j) = h_1(Y_i)$ is $\leq k \cdot \frac{1}{k^2} = 1/k$, and so the total expected communication cost for stage \ref{commstagethree} over all $i$ with $Y_i \notin X$ is $O(1)$.
Thus, the above protocol has the claimed total communication cost.
\end{proof}

In the following, we prove the communication lower bound.
First by Yao's Minimax Principle~\cite{yao1977probabilistic}, we may fix an input distribution and assume the protocol is deterministic.
Now let us consider the following hard distribution.

\paragraph{Hard distribution.}
We take the first $k^2$ elements from the universe, and divide them into $k$ blocks of size $k$ each.
$X$ will contain one uniformly random element from each block independently.
Each $Y_i$ will be a uniformly random element from the first $k^2$ elements.
Different $Y_i$'s are chosen independently.

The high-level idea of the proof is to first reduce from OSI to a classic (non-online) communication complexity
problem. 
In particular, we consider the problem solved in each stage of the OSI problem: Alice is given a set of $k$ elements from a universe of size $n$ and Bob
is given a single element from the same universe, and their goal is to determine if Bob's element is in Alice's set. 
This is precisely the \emph{index} problem.
Then we focus on the stage that costs the least amount of communication, and show an index lower bound with respect to this stage.
The hard distribution for OSI induces the following hard distribution for index.
\paragraph{Hard distribution for index.}
Divide the first $k^2$ elements of the universe into $k$ blocks of size $k$ each.
Alice's set $X$ consists of one uniformly random elements from each block independently.
Bob's element $y$ is chosen from the first $k^2$ elements uniformly at random.

We now prove a general lemma which relates protocols for ``symmetric'' online
communication problems (in which each round is essentially the same problem)
with protocols for classical communication problems. Note that when applied to
OSI, the associated single-round problem is index. In other words, a protocol
for OSI can be transformed into a very rigid protocol for index, which will be
easier for us to bound. Additionally, since we prove an iff statement, we know
that this transformation is lossless (up to constants).
\begin{lemma}\label{lem_symmetric_online_is_upfront}
  Suppose we have a problem in our online communication model and associated
  input distribution $\mathcal{D}$ over $\cX \times \cY^k$ with the following properties:
  \begin{enumerate}
    \item The function that Alice and Bob want to compute in each round,
      $f_i(X, Y_i)$, does not depend on the round number $i$.
    \item Conditioned on Alice's input $X \in \cX$, Bob's inputs
      $Y_1, \ldots, Y_k \in \cY$ are identically (but not necessarily
      independently) distributed.
  \end{enumerate}
  
  The associated single-round classical problem and associated input
  distribution are as follows. Alice is given an input $X \in \cX$ and Bob is
  given an input $Y \in \cY$, and they want to compute $f_1(X, Y)$. Their
  inputs are obtained by drawing an input $(X, Y_1, \ldots, Y_k)$ from
  $\mathcal{D}$, giving Alice $X$, and giving Bob $Y = Y_1$.
  
  There is a protocol for the online problem which uses $O(g(n, k))$ bits in
  expectation if and only if there is a protocol for the associated single-round
  problem where Alice first sends a message of expected length $O(g(n, k))$ bits
  and then Alice and Bob only speak an additional $O(g(n, k) / k)$ bits in
  expectation.
\end{lemma}

\begin{proof}
  We first prove the more nuanced forward direction. Suppose we have such a
  protocol $P$ for the online problem; we want a protocol $P'$ for the
  associated single-round problem with the above properties.
  
  The key idea is to focus on the stage where the players send the least bits in
  expectation. Choose $i \in [k]$ such that the players only speak
  $O(g(n, k) / k)$ bits in expectation in stage $i$. To solve the associated
  single-round problem on $(X, Y)$, we use the following protocol $P'$:
  \begin{itemize}
    \item The players pretend that they were given an online input where Alice
      received $X$ and Bob received $Y_i = Y$. They use public randomness to
      sample $Y_1, \ldots, Y_{i-1}$ according to $\mathcal{D}$.
    \item Alice has all the information for the first $i-1$ stages, so she
      simulates those stages of $P$ \emph{for both players}. Note this is
      possible because $P$ is an online protocol, and hence this simulation does
      not depend on any of $Y_i, \ldots, Y_k$. Alice then sends Bob the entire
      transcript.
    \item Alice and Bob then communicate to simulate stage $i$ of $P$,
      continuing from the transcript that Alice sent in the previous step.
    \item Bob outputs $P$'s decision about $f_i(X, Y_i)$.
  \end{itemize}
  
  In this protocol $P'$, the first message is sent by Alice in step (2). It has
  expected length no more than the transcript of $P$, which is $O(g(n, k))$.
  The players then simulate stage $i$ in step (3). Since the imaginary input
  follows distribution $\mathcal{D}$, the expected communication in this step is
  $O(g(n, k) / k)$. Since the goal of stage $i$ in the online problem is to
  compute $f_i(X, Y_i)$, which is precisely $f_1(X, Y)$ by our assumption
  about $f$ and choosing $Y_i = Y$. Hence the output of $P'$ is correct is $P$
  is correct.
  
  We finish with the easier reverse direction. Suppose we have such a protocol
  $P'$ for the associated single-round problem; we want a protocol $P$ for the
  online problem with the above properties.
  
  By construction, when following $P'$, Alice first sends a message with
  $O(g(n, k))$ bits in expectation. This message can only depend on her input.
  Our protocol $P$ also begins with Alice sending this message before Bob begins
  speaking. Now, in each stage, Bob is given an input $Y_i$. Alice and Bob
  can simulate $P'$ on $(X, Y_i)$, but skipping the initial message from Alice
  since it has already been sent.
  
  In our protocol $P$, Alice sends $O(g(n, k))$ bits in expectation in her first
  message. Then in each stage, only $O(g(n, k) / k)$ bits in expectation are
  transmitted between the players. Note that we just used the assumption that
  $Y_i$ and $Y_1$ are identically distributed conditioned on $X$; this is why
  $P'$ has the usual expected communication cost when run on $(X, Y_i)$. Thus
  the total communication cost is $O(g(n, k))$ bits in expectation.
\end{proof}

\ifx\undefined\odisj
\ifx\undefined\mainfile



\fi
\fi


Let $P'$ be a zero-error protocol for index such that Alice first sends $c_0$ bits in expectation, and then Alice and Bob communicate for $c_A$ and $c_B$ bits respectively (in expectation).
The following lemma lower bounds $c_0, c_A, c_B$.

\begin{lemma}\label{lem_index}
For sufficiently large $k$, any such $P'$ must have either
\begin{itemize}
	\item
		$c_0\geq \frac{1}{7}k\log k$, or
	\item
		$c_A\geq c_0\cdot 2^{-13\max\{c_B, 1\}\cdot 2^{6c_0/k}}$.
\end{itemize}

\end{lemma}

The main idea of the proof is to let Alice simulate Bob.
For simplicity, let us first assume the protocol has three rounds: Alice sends $c_0$ bits, then Bob sends $c_B$ bits, finally Alice sends $c_A$ bits.
To simulate Bob, Alice goes over all possible messages that Bob could send, then for each message, sends Bob what she would say if she received that message.
If Bob sends at most $c_B$ bits \emph{in worst case}, Alice will be able to complete the above simulate in $c_0+c_A\cdot 2^{c_B}$ bits of communication.
Then Bob will output whether his input $Y_i$ is in Alice's set $X$.
In particular, Alice's message depends only her input $X$, and Bob can do so for any $Y_i$.
That is, Bob will be able to recover the set $X$ based only on this message, which yields a lower bound on $c_0, c_A, c_B$.

\begin{proof}[Proof of Lemma~\ref{lem_index}]
Without loss of generality, we may first assume $c_B\geq 1$.
By Markov's inequality and a union bound, for any $C\geq 2$, with probability at least $1-2/C$, Alice sends no more than $C\cdot c_A$ bits and Bob sends no more than $C\cdot c_B$ bits after Alice's first message.
The next step is to let Alice \emph{simulate} the entire protocol, and turn it into \emph{one-way communication}.

More specifically, the transcript $\pi$ of a conversation between Alice and Bob is a binary string, in which each bit represents the message sent in the chronological order. 
Given $\pi$ and a fixed protocol, there shall be no ambiguity in which bits are sent by which player. 
That is, for any $\pi$, we can always decompose it into $(\pi_A, \pi_B)$, where $\pi_A$ is a binary string obtained by concatenating the bits sent by Alice in the chronological order, and similar for $\pi_B$. 
On the other hand, given $(\pi_A,\pi_B)$, there is a \emph{unique} way to combine them into a single transcript $\pi$, since a prefix of the transcript uniquely determines the player who speaks the next.
We know that with probability at least $1-2/C$, $|\pi_A|\leq C\cdot c_A$ and $|\pi_B|\leq C\cdot c_B$. 
In the new protocol, after Alice sends the first $c_0\cdot k$ bits, she goes over all $2^{C\cdot c_B}$ strings $s$ of length at most $C\cdot c_B$.
For each $s$ (in alphabetical order), she sends the first $C\cdot c_A$ bits of $\pi_A$ based on her input \emph{assuming $\pi_B=s$}.
That is, Alice tells Bob that ``if $s$ was your first $C\cdot c_B$ bits of the conversation, then here is what I would say for my first $C\cdot c_A$ bits.'' 
In total, she sends another $C\cdot c_A\cdot 2^{C\cdot c_B}$ bits. 
Thus, Bob can figure out the answer based only on the above messages, with probability $1-2/C$ (over the random input pairs).
To balance the lengths of two messages, we set $C=\frac{1}{2c_B}\log \frac{c_0}{c_A}$. 
If $C<2$, then we have $\log \frac{c_0}{c_A}<4c_B$, and thus
\[
	c_A>c_0\cdot 2^{-4c_B},
\]
which implies the second inequality in the statement.
Otherwise, the above argument holds, and we have 
\[
\begin{aligned}
C\cdot c_A\cdot 2^{C\cdot c_B}&=C\cdot c_A\cdot \sqrt{\frac{c_0}{c_A}} \\
&=\frac{c_A}{2c_B}\cdot \left(\sqrt{\frac{c_0}{c_A}}\log \frac{c_0}{c_A}\right)\\
&\leq c_A\cdot \left(\sqrt{\frac{c_0}{c_A}}\log \sqrt{\frac{c_0 }{c_A}}\right) \\
&\leq c_A\cdot \frac{c_0}{c_A}=c_0.
\end{aligned}
\]

Thus, Alice sends at most $2c_0$ bits in expectation in total. 
This message only depends on her input $X$. 
By Markov's inequality, for at least $2/3$ of the $X$'s, Alice sends no more than $6c_0$ bits. 
By Markov's inequality again, for at least $2/3$ of the $X$'s, the probability (over a random $y$) that Bob can figure out if $b\in A$ based only on Alice's first message is at least $6/C$. 
Since there are $k^k$ different possible $X$'s, at least $k^k/3$ different $X$'s have both conditions hold. 
Thus, there must be $k^k/3\cdot 2^{-6c_0
}$ such $X$'s that Alice sends the same message $M$.
Denote this set of $X$'s by $\cX$.
Moreover, when $M$ is the message Bob receives, there are at least $(1-6/C)k^2$ different $y$'s such that Bob can figure out the answer based only on the value of $y$ and $M$.
Denote this set of $y$'s by $\cY$.
In the combinatorial rectangle $\cR=\cX\times \cY$, for every $y\in \cY$, either $y\in X$ for every $X\in\cX$, or $y\notin X$ for every $X\in\cX$.
That is, $\mathcal{R}$ is a \emph{column-monochromatic rectangle}\footnote{A rectangle with the same function value in every column.} of size $(k^k/3\cdot 2^{-6c_0
})\times (1-6/C)k^2$.

On the other hand, for the index problem, in any column-monochromatic rectangle $\mathcal{R}=\mathcal{X}\times \mathcal{Y}$, the answer is ``YES'' in no more than $k$ columns of $\mathcal{Y}$ (the element is in the set). 
This is because each set $X\in\cX$ has size $k$. 
In order to upper bound the number of $y\in\cY$ that is not in any $X$, let $r_i$ for $1\leq i\leq k$ be the size of the intersection of $\cY$ and the $i$-th block of the universe.
Thus, the number of $X$'s that avoids all $y\in\cY$ is at most
\[
	(k-r_1)(k-r_2) \cdots (k-r_k)\leq \left(k-\frac{1}{k}(r_1+\cdots+r_k)\right)^k
\]
by the AM-GM inequality. 
That is, at most $k^2-k|\cX|^{1/k}$ $y$ are not in any $X$.
Overall, we have $|Y|\leq k+k^2-k|\cX|^{1/k}$.
Combining this with the parameters from the last paragraph, we get
\[
	(1-6/C)k^2\leq k+k^2-k\left((k^k/3\cdot 2^{-6c_0})\right)^{1/k}.
\]
Simplifying the inequality yields
\[
	6/C\geq 2^{-6c_0/k}\cdot 3^{-1/k}-1/k.
\]
When $c_0<\frac{1}{7}k\log k$, we have $2^{-6c_0/k}\cdot 3^{-1/k}-1/k>\frac{12}{13}\cdot 2^{-6c_0/k}$ for sufficiently large $k$.
Pluging-in the value of $C(=\frac{1}{2c_B}\log \frac{c_0}{c_A})$ and simplifying, we obtain
\[
c_A\geq c_0\cdot 2^{-13c_B/2^{-6c_0/k}}.
\]
This proves the lemma.
\end{proof}
\ifx\undefined\odisj
\ifx\undefined\mainfile
\end{document}

\fi
\fi

\begin{proof}[Proof of Theorem~\ref{thm_OSI_LB}]
For any OSI protocol with total communication cost $c$, by Lemma~\ref{lem_symmetric_online_is_upfront} and Lemma~\ref{lem_index}, we have either
\begin{itemize}
  \item
    $c\geq \frac{1}{7}k\log k$, or
  \item
    $c/k\geq c\cdot 2^{-13\max\{c/k,1\}\cdot 2^{6c/k}}$.
\end{itemize}
The second inequality simplifies to $\max\{c/k,1\}\cdot 2^{\Theta(c/k)}\geq \Omega(\log k)$.
Thus, we must have $c\geq \Omega(k\log\log k)$.
\end{proof}

\ifx\undefined\mainfile

\bibliographystyle{alpha}
\bibliography{online_disj}

\appendix

\end{document}
\fi

\ifx\undefined\mainfile


\fi

\section{The Group Range Problem}
\label{s:group-range-section}
\label{s:group-range}

The goal of this section is to prove our main result:
\begin{restate}[Theorem~\ref{thm:group_range_LB}]
There exists a distribution over $n$ updates and queries for the Group Range
Problem with binary encoding of the group elements $e:G\rightarrow \{0,1\}^s$,
such that for any randomized cell-probe data structure $D$ with word size
$w = \Theta(\log n)$, which with probability $p$ answers at least a
$(\frac{1}{2} + \delta)$ fraction of queries correctly and spends $\epsilon n \log n$
total running time, we must have $p \leq \exp(-\delta^2 n)$, as long as
$s \leq (1+\epsilon)\log |G|$, $\delta^2\gg\epsilon\geq \Omega(1/\log n)$, and
$n$ is sufficiently large.
\end{restate}

For convenience, we will assume that $n$ is a power
of two. 
A similar argument applies to the general case.
We will also say that the data structure succeeds on an input when the
event described occurs: it answers a $(\frac{1}{2} + \delta)$ fraction of
queries correctly and spends at most $\epsilon n \log n$ total running time.

Our proof is divided into three steps. First, we construct a random input
sequence so that we can apply Yao's minimax principle and consider a
determinstic data structure. Second, we consider various subproblems of this
sequence. We show that the data structure must do well on at least one of them,
but with some additional structure on how it probes cells when solving this
subproblem. Third, we use the data structure on this subproblem to produce a
communication protocol for a problem in our online communication complexity
model.

\subsection{Step One: The Hard Distribution}

\begin{figure}
  \centering
  \begin{tabular}{| c | c |}
    \hline
    \multicolumn{1}{| c |}{\bfseries Shorthand} & \multicolumn{1}{| c |}{\bfseries Operation} \\ \hline
    $u_0$ & Update($0, \unif_G$) \\ \hline
    $q_0$ & Query($0, \unif_{[n]}, \unif_{[s]}$) \\ \hline
    $u_1$ & Update($2, \unif_G$) \\ \hline
    $q_1$ & Query($0, \unif_{[n]}, \unif_{[s]}$) \\ \hline
    $u_2$ & Update($1, \unif_G$) \\ \hline
    $q_2$ & Query($0, \unif_{[n]}, \unif_{[s]}$) \\ \hline
    $u_3$ & Update($3, \unif_G$) \\ \hline
    $q_4$ & Query($0, \unif_{[n]}, \unif_{[s]}$) \\ \hline
  \end{tabular}
\caption{Structure of our random input sequence ($n = 4$). Here, $\unif_S$ is
  an entry drawn from the uniform distribution on set $S$.}
\label{f:input}
\end{figure}

Our random input sequence for $D$ has the following essential properties:
\begin{enumerate}
  \item[(i)] Update and query operations are interleaved.
  \item[(ii)] If we look at any contiguous window of operations, the update operations
    are always somewhat spread out over all elements.
  \item[(iii)] Each query operation checks a random prefix of the sequence.
\end{enumerate}

Regarding the first property, our sequence consists of $2n$ alternating update and
query operations:\\ $(u_0, q_0, u_1, q_1, \ldots, u_{n-1}, q_{n-1})$.

Next, we define the update operations $u_i$, keeping property $(ii)$ in mind.
For this, we use a standard trick: the reversed binary representation. Let
$\rev{s}{\cdot}$ reverse $s$-bit integers, e.g. $\rev{s}{1} = 2^{s-1}$ and
$\rev{s}{2^s - 1} = 2^s - 1$. The $i^{th}$ update operation sets group element
$\rev{s}{i}$ to a uniform random group element.

We finish by defining the query operations $q_i$. For each query operation, we
need a range and the index of a bit. Our range will be $[0, R]$, where $R$ is
drawn uniformly from $\{0, 1, \ldots, n-1\}$. The bit index will be selected
uniformly at random over all indices.

Figure~\ref{f:input} shows what a random input looks like in the $n = 4$ case.
Yao's minimax principle guarantees that since $D$ is a randomized structure with
guarantees on worst-case inputs, there must be a deterministic data structure
$D'$ with the same guarantees on a random input sequence of this form.

\ifx\undefined\mainfile
\end{document}
\fi
\ifx\undefined\mainfile


\fi

\subsection{Step Two: Identifying Key Subproblems}

In this section, we give the formal details on how to identify key subproblems of a data structure problem that we will be able to later transform into online communication games. We begin by describing these subproblems, and later prove several key properties about them. Suppose we have a data structure problem which involves updates (operations which produce no output) and queries (operations which produce output). We also have a hard input distribution, which produces input sequences consisting of $n$ \emph{operation blocks}. Each block contains $n_b$ operations. Hence, input sequences have $N = n \cdot n_b$ operations in total.

\begin{figure}
\centering
\begin{tabular}{| c | c | c | c | c | c | c | c | c |}
  \hline
  Blocks & $B_0$ & $B_1$ & $B_2$ & $B_3$ & $B_4$ & $B_5$ & $B_6$ & $B_7$ \\ \hline
  Level 0 & \multicolumn{4}{| c |}{$I_A(v_1)$} & \multicolumn{4}{| c |}{$I_B(v_1)$} \\ \hline
  Level 1 & \multicolumn{2}{| c |}{$I_A(v_2)$} & \multicolumn{2}{| c |}{$I_B(v_2)$} &
    \multicolumn{2}{| c |}{$I_A(v_3)$} & \multicolumn{2}{| c |}{$I_B(v_3)$} \\ \hline
  Level 2 & $I_A(v_4)$ & $I_B(v_4)$ &
    $I_A(v_5)$ & $I_B(v_5)$ &
    $I_A(v_6)$ & $I_B(v_6)$ &
    $I_A(v_7)$ & $I_B(v_7)$ \\ \hline
\end{tabular}
\caption{Division into subproblem intervals ($n = 8$).}
\label{f:subproblem-intervals}
\end{figure}

As shown in Figure~\ref{f:subproblem-intervals}, each of our subproblems designates two equally-sized adjacent intervals of blocks. The earlier interval is $I_A$ (Alice's interval), and the later interval is $I_B$ (Bob's interval). In the first subproblem, Alice's interval is the first half of the input sequence and Bob's interval is the second half. In the second subproblem, Alice's interval is the first quarter; Bob's interval, the second quarter. Roughly speaking, we keep recursively dividing Alice's interval and Bob's interval to get smaller subproblems. Note that our subproblems overlap quite a bit; each phase can be found in $\Theta(\log n)$ subproblems.

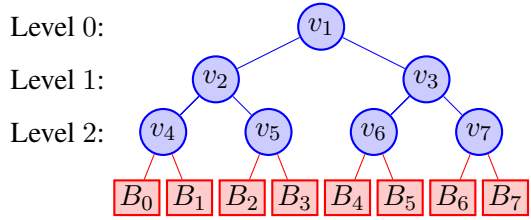
\begin{figure}
\centering
\begin{tikzpicture}[%
  auto,
  scale=0.35,
  vertex/.style={
    circle,
    draw=blue,
    thick,
    fill=blue!20,
    align=center,
    inner sep=2pt
  },
  leaf/.style={
    rectangle,
    draw=red,
    thick,
    fill=red!20,
    align=center,
    inner sep=2pt
  }
  ]
  \node[vertex] (v_1) at (5, 5) {$v_1$};
  \node[vertex] (v_2) at (1, 3) {$v_2$};
  \node[vertex] (v_3) at (9, 3) {$v_3$};
  \node[vertex] (v_4) at (-1, 1) {$v_4$};
  \node[vertex] (v_5) at (3, 1) {$v_5$};
  \node[vertex] (v_6) at (7, 1) {$v_6$};
  \node[vertex] (v_7) at (11, 1) {$v_7$};

  \foreach \i in {0,1,...,7}
    \node[leaf] (l_\i) at (\i*2-2, -1.5) {$B_\i$};
  
  \node (level0) at (-5, 5) {Level $0$:};
  \node (level1) at (-5, 3) {Level $1$:};
  \node (level2) at (-5, 1) {Level $2$:};
  
  \draw[blue] (v_1) -- (v_2) -- (v_4) -- (v_2) -- (v_5);
  \draw[blue] (v_1) -- (v_3) -- (v_6) -- (v_3) -- (v_7);

  \foreach \i/\j in {4/0,4/1,5/2,5/3,6/4,6/5,7/6,7/7}
    \draw[red] (v_\i) -- (l_\j);
\end{tikzpicture}
\caption{Our subproblems correspond to nodes of a balanced binary tree ($n = 8$).}
\label{f:subproblem-tree}
\end{figure}

To be more formal, consider a balanced binary tree with $(\log_2 n + 1)$ levels (depicted in Figure~\ref{f:subproblem-tree}). The operation blocks are the leaves of this tree. Such a tree has $n-1$ \emph{internal} nodes: $v_1, \ldots, v_{n-1}$, where $v_1$ is the root node and the children of node $v_i$ are nodes $v_{2i}$ (left) and $v_{2i + 1}$ (right). Consider the subtree rooted at $v$; the leaves of this subtree form a contiguous interval of operation blocks. We denote the first (left) half of this interval as $I_A(v)$, and the second (right) half as $I_B(v)$. Each internal node $v$ corresponds to a subproblem, which we denote with the interval pair $(I_A(v), I_B(v))$. For each subproblem, we are interested in the number of memory cells that the data structure probes at least once in $I_A$, and at least once in $I_B$. Intuitively speaking, this captures the amount of information being stored about updates which is later used to answer queries. We need some additional notation to discuss this tree and the cells being probed:
\begin{definition}
  The set of nodes in level $j$ is denoted $\ell(j)$ and consists of $\{ v_{2^j}, \ldots, v_{2^{j+1}-1} \}$.
  The set of cells that the data structure probes when processing the phases of $I_A(v)$ is $P_A(v)$ (P stands for probes). Similarly, the set of probed cells when processing $I_B(v)$ is denoted $P_B(v)$.
\end{definition}

We restate our focus using this new notation: for each subproblem, we are interested in the value of $|P_A(v) \cap P_B(v)|$. We now state the general reduction that we aim to prove in this subsection.
\begin{theorem}\label{thm:step2}
  Suppose that there is a data structure problem along with a hard distribution for it over sequences of $n$ blocks consisting of $n_b$ operations each, for a total of $N = n n_b$ operations. Next, suppose there exists a constant $c \in (0, 1)$, value $\epsilon_0>0$, and a bivariate convex function $g(x, y)$, whose value is non-decreasing in $x$ and non-increasing in $y$, so that the following is true: for any data structure $D$, any subproblem $(I_A(v), I_B(v))$ where $I_B(v)$ consists of $k \ge n^{1-c}$ blocks, any $\epsilon_v \geq 0$ and $\delta_v\in [0,1/2]$, the probability \emph{conditioned on} all operations before $I_A(v)$ that the following hold:
  \begin{itemize}
    \item $\SharedProbes \le \epsilon_v \cdot k n_b$,
    \item $D$ answers a $(\frac12 + \delta_v)$-fraction of queries in $I_B(v)$ correctly,
  \end{itemize}
  is at most $\exp \left( -g(\delta_v,\epsilon_v) k n_b \right)$.
  Then the probability that all the following hold:
  \begin{itemize}
    \item $D$ probes at most $\epsilon N \log n$ cells,
    \item $D$ answers a $(\frac12 + \delta)$-fraction of all queries correctly,
  \end{itemize}
  is at most $\exp(n^c\cdot \log N) \cdot \exp\left(-g(\delta-3/\sqrt{c\log n}, \epsilon/c) \cdot N\right)$ as long as $\delta\geq 3/\sqrt{c\log n}$.
\end{theorem}

\begin{proof}
For convenience, when the data structure meets the first set of conditions for a subproblem (i.e., efficient and accurate for this subproblem), we will say that it ``succeeds'' at the subproblem. Similarly, when it meets the second set of conditions for an input (i.e., efficient and accurate overall), we will say that it ``succeeds'' on that input.

In this proof, we need to maintain both efficiency and accuracy guarantees when identifying the right subproblems.
We begin by explaining the efficiency conditions and their relation to total cell probes.

Consider the sum $\sum_{j \in \AllLevels} \sum_{v \in \ell(j)} \SharedProbes$. Each time the data structure probes a cell, it contributes to at most a single term in this summation: the one where its previous access to the cell was in $P_A(v)$ and its current access is in $P_B(v)$. Hence this sum is upper bounded by the total number of cell probes. When the data structure succeeds on an input, we know that:
\[
  \sum_{j \in [\log n]} \sum_{v \in \ell(j)} |P_A(v) \cap P_B(v)| \le \epsilon N \log n
\]
in addition to correctly answering a $(\frac12 + \delta)$ fraction of all queries.

Our plan of attack is to apply the first set of conditions for a subproblem to all subproblems in an entire level. Our first task is to identify the right level. In addition to this level being efficient and accurate, it also cannot be in the the bottom of the tree, since our assumption requires $k \ge n^{1-c}$. The following definitions will help us in the process of identification:
\begin{definition}
  Let $\delta_j$ be a random variable so that a $(\frac12 + \delta_j)$ fraction of the queries in $\cup_{v \in \ell(j)} I_B(v)$ are answered correctly.
  Let $\epsilon_j$ be a random variable so that $\sum_{v \in \ell(j)} |P_A(v) \cap P_B(v)| = \epsilon_j N$.
\end{definition}

Therefore, the above inequality translates to 
\begin{equation}\label{eqn:eps_sum}
  \sum_{j\in[\log n]} \epsilon_j\leq \epsilon \log n.
\end{equation}

At first blush, it may seem that if our data structure manages to answer many queries correctly, it must do so over all intervals $I_B(v)$. Unfortunately, the matter is not so simple. Some of our queries may be duplicated a logarithmic number of times over different intervals $I_B(v)$, while others only show up once. For example, in Figure~\ref{f:subproblem-intervals}, queries in $B_0$ do not show up in any subproblems, while queries in $B_7$ show up in three subproblems. As our binary tree helpfully suggests, the secret to this behavior lies in the binary representation. Suppose we have a query in $B_i$; if we write $i$ as a binary number, we know the path we need to walk down the tree to get to $B_i$; we walk left at level $j$ if the $j^{th}$ bit (starting from most-significant) is zero, and right if it is one. But walking left or right exactly dictates whether $B_i$ will be in $I_A(v)$ or $I_B(v)$.

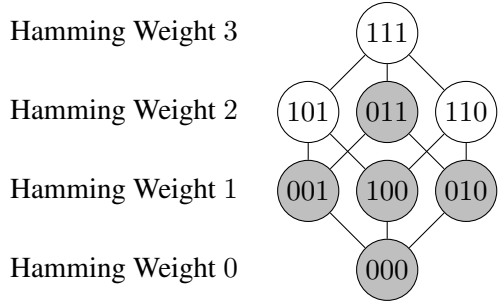
\begin{figure}
\centering
\begin{tikzpicture}[%
  auto,
  scale=0.35,
  hollownode/.style={
    circle,
    draw=black,
    inner sep=2pt
  },
  shadednode/.style={
    circle,
    draw=black,
    fill=black!25,
    inner sep=2pt
  },
  ]

  \node[shadednode] (1) at (0, 0) {$000$};
  \node[shadednode] (2) at (-3, 3) {$001$};
  \node[shadednode] (3) at (3, 3) {$010$};
  \node[shadednode] (4) at (0, 6) {$011$};
  \node[shadednode] (5) at (0, 3) {$100$};
  \node[hollownode] (6) at (-3, 6) {$101$};
  \node[hollownode] (7) at (3, 6) {$110$};
  \node[hollownode] (8) at (0, 9) {$111$};

  \draw (1) -- (2) -- (4) -- (3) -- (1);
  \draw (5) -- (6) -- (8) -- (7) -- (5);
  \draw (1) -- (5);
  \draw (2) -- (6);
  \draw (3) -- (7);
  \draw (4) -- (8);
  
  \node (level0) at (-10, 0) {Hamming Weight $0$};
  \node (level1) at (-10, 3) {Hamming Weight $1$};
  \node (level2) at (-10, 6) {Hamming Weight $2$};
  \node (level3) at (-10, 9) {Hamming Weight $3$};
\end{tikzpicture}
\caption{Blocks lie at vertices of a hypercube. In this example, $n = 8$ and
there is only one query per block. Queries answered correctly are shaded, and
incorrect queries are not shaded. Even though $\frac58 = 62.5\%$ of queries in
the overall hypercube are correct, the average small subhypercube $H_j$ only has
$\frac{1+2+2}{12} = 41.\bar{6}\%$ of its queries correct, and no small hypercube
$H_j$ has more than $\frac24 = 50\%$ of its queries correct.}
\label{f:query-cube}
\end{figure}

The takeaway is that we can visualize our blocks as vertices of the $(d = \log n)$-dimensional Boolean hypercube $H$. The Hamming weight of a block tells us how many levels it appears in the $I_B$ of. If we look at all $I_B(v)$'s for all $v$ in level $j$, we get a lower-dimensional hypercube $H_j = \{x \in H \mid x_j = 1\}$. Suppose we label every vertex of the big hypercube $H$ (which corresponds to a block) with the fraction of queries answered correctly. We want to show that an average small hypercube $H_j$ still has a reasonable fraction of correct queries. Figure~\ref{f:query-cube} depicts the situation and demonstrates that the fraction may decrease. We aim to bound this loss.

Since a node of Hamming weight $w$ contributes to exactly $w$ subhypercubes, the worst case labelling occurs when nodes with the lowest Hamming weight are assigned correct queries first. Suppose $(\frac12+\delta)$-fraction of nodes have Hamming weight at most $W$, i.e.,
let $W$ be the largest integer such that
\[
  \sum_{w=0}^W {d \choose w} \le \left( \frac12+\delta \right) 2^d.
\]
Therefore, we have $\sum_{w=\lceil \frac d2 \rceil+1}^W {d\choose w} \geq (\delta - 3/\sqrt{d}) 2^d$. Then the average fraction of correct queries in a random subhypercube is at least:
\begin{align*}
  \frac1d \sum_{w=0}^W {d\choose w}\cdot w\cdot 2^{-(d-1)}
    &=\frac1d \sum_{w=0}^{\lceil d/2\rceil} {d\choose w}\cdot w\cdot 2^{-(d-1)}
      + \frac1d \sum_{w=\lceil d/2\rceil+1}^W {d\choose w}\cdot w\cdot 2^{-(d-1)} \\
    &\ge \sum_{w=1}^{\lceil d/2\rceil} {d-1\choose w-1}\cdot 2^{-(d-1)}
      + \frac{1}{d}\sum_{w=\lceil d/2\rceil+1}^W {d\choose w}\cdot \frac{d}{2}\cdot 2^{-(d-1)} \\
    &\ge \frac12 + \sum_{w=\lceil d/2\rceil+1}^W {d\choose w}\cdot 2^{-d} \\
    &\ge \frac12 + \delta-3/\sqrt{\log n}.
\end{align*}

We have managed to get a bound on the accuracy over levels:
\[
  \avg_{j \in \AllLevels} \delta_j \geq \delta - 3/\sqrt{\log n},
\]
where $\avg_{j\in S} a_j := \frac{1}{|S|}\sum_{j\in S} a_j$ is the average value.
However, recall that we want to avoid the bottom levels of the tree, since we need $k \ge n^{1-c}$. Hence we restrict to the top $L = c \log n$ levels. 
By applying the same proof to only top $L$ levels, we have
\[
  \avg_{j \in \TopLevels} \delta_j \geq \delta - 3/\sqrt{c\log n}.
\]

We now have the accuracy half of our task of identifying the right level. We want to wind up showing that when the data structure succeeds on its input, there is some level $j \in \TopLevels$ such that $g(\delta_j,\epsilon_j)$ is large.

By Equation~\eqref{eqn:eps_sum}, we have $\avg_{j \in [L]} \epsilon_j \le \epsilon/c$ . Hence by Jensen's inequality, we know that 
\begin{align*}
  \avg_{j\in [L]} g(\delta_j, \epsilon_j)
    &\ge g(\avg_{j\in [L]}\delta_j, \avg_{j\in [L]}\epsilon_j) \\
    &\ge g(\delta-3/\sqrt{c\log n}, \epsilon/c),
\end{align*}
and therefore, there exists some level $j$ with large $g(\delta_j,\epsilon_j)$ value:
\[
  \max_{j \in \TopLevels} g(\delta_j, \epsilon_j)\ge g(\delta-3/\sqrt{c\log n}, \epsilon/c).
\]
Note that in the above inequality, we used the convexity and monotonicity of $f$.

Whenever our data structure succeeds on its input, there must be some level $j \in \TopLevels$ with this guarantee on $\delta_j$ and $\epsilon_j$. 
Hence to upper bound its probability, it suffices to prove for every level $j\in\TopLevels$, the probability that $g(\delta_j,\epsilon_j)$ is large is tiny.
Then an application of union bound over all levels $j$ would prove the theorem.

We have finished identifying our level of interest, and want to repeatedly apply the first set of conditions to all of its subproblems. Analogous to how we defined accuracy parameters $\delta_j$ and efficiency parameters $\epsilon_j$ for levels, we can define these parameters for each subproblem:
\begin{definition}
  Let $\delta_v$ be a random variable so that a $(\frac12 + \delta_v)$ fraction of the queries in $I_B(v)$ are answered correctly.
  Let $\epsilon_v$ be a random variable so that $\SharedProbes = \epsilon_v k n_b$ where $k$ is the number of operation blocks in $I_B(v)$.
\end{definition}

Fix two sequences $\{\delta_v\}_{v \in \ell(j)}$ and $\{\epsilon_v\}_{v\in \ell(j)}$ with the correct averages: $\sum_{v \in \ell(j)} \delta_v = \frac{n}{k} \delta_j$ and $\sum_{v \in \ell(j)} \epsilon_v = \frac{n}{k} \epsilon_j$. We will apply the hypothesis to each subproblem $v \in \ell(j)$ with good parameters $(\delta_v, \epsilon_v)$ and $\epsilon_v$. By noticing that all the interval pairs $(I_A(v), I_B(v))$ are
disjoint, $k\geq n^{1-c}$, and by using Jensen's inequality, we have the probability that for all $v \in \ell(j)$:
\begin{itemize}
  \item $\SharedProbes \leq \epsilon_v\cdot k n_b$ and
  \item $D$ answers a $(1/2 + \delta_v)$-fraction of queries in $I_B(v)$ correctly
\end{itemize}
is at most:
\begin{align*}
  \prod_{v\in\ell(j)} \exp(-g(\delta_v, \epsilon_v)\cdot k n_b)
    &= \exp\left(-\sum_{v\in\ell(j)}g(\delta_v, \epsilon_v)\cdot k n_b\right) \\
    &= \exp\left(-\avg_{v\in\ell(j)}g(\delta_v, \epsilon_v)\cdot n n_b\right) \\
    &\le \exp\left(-g\left(\avg_{v\in\ell(n)}\delta_v, \avg_{v\in\ell(n)}\epsilon_v\right) N\right)\\
    &\leq \exp\left(-g(\delta-3/\sqrt{c\log n}, \epsilon/c) \cdot N\right).
\end{align*}

We can finish by taking a union bound over all possible sequences $\{\delta_v\}_{v \in \ell(j)}$ and $\{\epsilon_v\}_{v\in \ell(j)}$. There are at most ${N+n/k \choose n/k}$ possibilities for the first sequence and ${\epsilon_j N + n/k \choose n/k}$ for the second sequence, so the probability of level $j$ having good guarantees can be at most:
\begin{align*}
  & {N+n/k\choose n/k} \cdot {\epsilon_j N+n/k\choose n/k} \cdot \exp\left(-g(\delta-3/\sqrt{c\log n}, \epsilon/c) \cdot N\right) \\
  &\le \exp(n/k\cdot \log k n_b) \cdot \exp\left(-g(\delta-3/\sqrt{c\log n}, \epsilon/c) \cdot N\right) \\
  &\le \exp(n^c\cdot \log N) \cdot \exp\left(-g(\delta-3/\sqrt{c\log n}, \epsilon/c) \cdot N\right)
\end{align*}
The last inequality holds because our $k$ was at least $n^{1-c}$.
This completes the proof of Theorem~\ref{thm:step2}.
\end{proof}

\ifx\undefined\mainfile
\end{document}
\fi
\ifx\undefined\mainfile


\fi

We now relate the lower bound on subproblems that we will prove with our desired data structure lower bound, Theorem~\ref{thm:group_range_LB}. We will need the following lemma about subproblems, whose proof is deferred to the next subsection.
\begin{lemma}\label{lem:interval_pair}
  Suppose we have two intervals $I_A = I_A(v)$ and $I_B = I_B(v)$ consisting of
  $k$ updates and queries each. Then the probability that
  \begin{itemize}
    \item
      $\left|P_A(v)\cap P_B(v)\right|\leq \epsilon_v \cdot k$ and
    \item
      $D$ answers a $(1/2+\delta_v)$-fraction of queries in $I_B(v)$ correctly
  \end{itemize}
  conditioned on all operations $O$ before $I_A(v)$ is at most
  \[
  \exp(-(\delta_v-\beta\cdot (\sqrt{\epsilon}+\sqrt{\epsilon_v}))^2\cdot k)
  \]
  for some constant $\beta>0$, as long as $s \le (1 + \epsilon) \log |G|$, $k\gg\log n$, and $\delta_v-\beta\cdot (\sqrt{\epsilon}+\sqrt{\epsilon_v})\geq 0$.
\end{lemma}

Since Theorem~\ref{thm:step2} has a convexity requirement, we will also need the following technical lemma about the convexity of our error function:
\begin{lemma}\label{lem_group_range_convex}
  For any $\beta > 0$ and $\epsilon \ge 0$, the function $g(x, y) = (\max \{0, x - \beta \sqrt{y} - \beta \sqrt{\epsilon} \})^2$ is convex over $(x, y) \in [0, \infty) \times [0, \infty)$.
\end{lemma}

\begin{proof}
  First, we write $g(x, y) = g_1(g_2(x, y))$, where $g_1(z) = z^2$ and $g_2(x, y) = \max \{0, x - \beta \sqrt{y} - \beta \sqrt{\epsilon} \}$. We will prove that $g_2(x, y)$ is convex over $(x, y) \in [0, \infty) \times [0, \infty)$. Combining that with the fact that $g_1$ is convex and nondecreasing over $z \in [0, \infty)$, and that the output of $g_2$ is always nonnegative, we will get that $g$ is convex.
  
  The max of two convex functions is convex. $g_3(x, y) = 0$ is a constant function, so it is convex. It suffices to prove that $g_4(x, y) = x - \beta \sqrt{y} - \beta \sqrt{\epsilon}$ is convex as well. We do so by showing that it is the sum of two convex functions: $g_5(x, y) = x - \beta \sqrt{\epsilon}$ and $g_6(x, y) = - \beta \sqrt{y}$. The former is linear, so it is convex.
  
  We compute the second derivative of $g_6$, since it only depends on a single variable.
  \begin{align*}
    g_6(y) &= - \beta \sqrt{y} \\
    g_6'(y) &= -\frac{\beta}{2} y^{-1/2} \\
    g_6''(y) &= \frac{\beta}{4} y^{-3/2}
  \end{align*}
  Hence for nonnegative $y$, $g_6$ is convex. Combining our convexity claims, the original function $g$ is convex over the desired range.
\end{proof}

We now have the tools necessary to prove Theorem~\ref{thm:group_range_LB}.

\begin{proof}[Proof of Theorem~\ref{thm:group_range_LB}]
  By Lemma~\ref{lem_group_range_convex}, we know that our error function $g$ is convex when $\delta_v, \epsilon_v \ge 0$. Hence we can invoke Theorem~\ref{thm:step2} with $c = 1/2$, since we only need that $k \gg \log n$. Note that Lemma~\ref{lem:interval_pair} required that $\delta_v-\beta\cdot (\sqrt{\epsilon}+\sqrt{\epsilon_v})\geq 0$, but when this is not true we can always use the trivial probability bound of $1 = e^0$ (this is why our error function $g$ has a $\max \{0, \cdot \}$). Also, note both the subproblem lower bound and resulting data structure lower bound share the same bound on $s$.
  
  This theorem invocation tells us that the probability that the data structure succeeds is at most:
  \begin{align*}
    &\exp(\sqrt{n} \log N) \cdot
      \exp (-\left(\max \{0, \delta - 3/\sqrt{0.5 \log n} - \beta\sqrt{2\epsilon} - \beta \sqrt{\epsilon} \} \right)^2 \cdot N) \\
    &\le \exp(\sqrt{n} \log N) \cdot \exp (-\left(\delta - 3/\sqrt{0.5 \log n} - \beta \sqrt{2\epsilon} - \beta \sqrt{\epsilon} \right)^2 \cdot N)
  \end{align*}
  
  The first exponential is dominated by the second. Since we assumed $\delta \gg \sqrt{\epsilon}$ and $\delta \gg \Omega(1 / \sqrt{\log n})$, the second exponential is simply $\exp(-\delta^2 \cdot N)$, completing the proof.
\end{proof}

\subsection{Step Three: The Communication Game}
\label{sec_step3}





The main goal of this step is to prove Lemma~\ref{lem:interval_pair}. The key
idea is to show how an efficient data structure can be used to produce an efficient communication
protocol for a particular communication game, and then to rule out the possibility of an efficient communication
protocol, hence proving that the original efficient data structure could not exist.
We begin by defining the communication game on interval pairs we will be focusing on, which uses our online
communication model from Section~\ref{sec:model}.

\paragraph*{Communication Game}
We define one communication game for each interval pair $(I_A, I_B)$.
Fix two intervals $I_A = I_A(v)$ and $I_B = I_B(v)$ consisting of $k$ updates and
queries each, all operations $O$ prior to these intervals, all queries $Q_A$ in
$I_A$ and all updates $U_B$ in $I_B$. That is, the only undetermined operations
up to the end of $I_B$ are the updates in $I_A$ and the queries in $I_B$. We
embed these operations into a communication game. In the associated online
communication game $G = G(v, O, Q_A, U_B)$, $X$ consists of the updates in $I_A$,
and $Y_i$ is the $i^{th}$ query in $I_B$. The goal of Stage $i$ is to compute
the $i^{th}$ query in $I_B$.\footnote{Note that the previous queries do not
affect the output of the $i^{th}$ query.}

\paragraph*{Input Distribution}
The input $X$ is sampled as a random set of updates in $I_A(v)$ and
$(Y_1, \ldots, Y_k)$ as a random set of queries in $I_B(v)$ under our hard
distribution for the Group Range problem. 

\begin{lemma}\label{lem:oc_upper}
Consider two intervals $I_A(v)$ and $I_B(v)$, consisting of $k$ updates and
queries each. Let the operations prior to them be $O$, the queries in $I_A(v)$ be
$Q_A$, the updates in $I_B(v)$ be $U_B$. For any data structure $D$ and $\epsilon_v$, there is a protocol $\cP_D$ for the communication game
$G(v, O, Q_A, U_B)$ such that
\begin{enumerate}
	\item
		Alice sends $2\epsilon_v\cdot k w+1$ bits;
	\item
		Bob sends no message;
	\item
		For every $\delta_v$, the probability that $\cP_D$ answers $(1/2+\delta_v-\epsilon_v)$-fraction of the $f_i(X,Y_i)$'s correctly is at least 
		\[
			\Pr\left[\left|P_A\cap P_B\right|\leq \epsilon_v\cdot k,\textrm{$D$ answers a } (1/2+\delta_v)\textrm{-fraction of queries in $I_B$ correctly}\,\,\middle|\,\, O,Q_A,U_B\right].
		\]
\end{enumerate}
\end{lemma}

\begin{lemma}\label{lem:oc_lower}
For any protocol $P$ for $G(v, O,Q_A,U_B)$ and $\epsilon_v,\delta_v$ where:
\begin{enumerate}
	\item
		Alice sends $O(\epsilon_v k\cdot \log n)$ bits, and
	\item
		Bob sends no message
\end{enumerate}
must have
\[
	\Pr[P\textrm{ answers $(1/2+\delta_v-\epsilon_v)$-fraction of the $f_i(X, Y_i)$'s correctly}]\leq\exp(-(\delta_v-O(\sqrt{\epsilon}+\sqrt{\epsilon_v}))^2\cdot k).
\]
\end{lemma}

Lemma~\ref{lem:interval_pair} follows directly from applying both Lemma~\ref{lem:oc_upper} and Lemma~\ref{lem:oc_lower}.
Hence it remains to prove these two lemmas.

\begin{proof}[Proof of Lemma~\ref{lem:oc_upper}]
The idea is that the players simulate $D$ as operations are revealed, and Alice
sends some necessary information to Bob. Consider the following protocol
$\cP_D$:
\begin{enumerate}
	\item (Preprocessing)
		Recall that Alice knows all operations up to the end of $I_A$ and the updates in $I_B$,
		Bob knows all operations prior to $I_A$ and all operations in $I_B$.
		First, Alice simulates $D$ up to the end of $I_A$, and Bob simulates $D$ up to the beginning of $I_A$ and \emph{skips $I_A$}.
		Denote the memory state that Alice has \emph{at this moment} by $M_A$.
		Next, the players are going to simulate operations in $I_B$.
	\item (Stage $i$ - Alice's simulation)
		Since the $(i-1)$-th query is revealed to Alice in the last stage, Alice continues the simulation up to the $i$-th query.
		Alice sends Bob the cells (their addresses and contents in $M_A$) that are
		\begin{itemize}
			\item
				probed during this part of the simulation, and
			\item
				probed during $I_A$, and
			\item
				not probed in the previous stages.
		\end{itemize}
		That is, Alice sends Bob all cells in $P_A\cap P_B$ that are just probed for the very first time among all stages so far.
	\item (Stage $i$ - Bob's simulation)
		Bob first updates his memory state according to Alice's message: For each cell in the message, Bob replaces its content with the actual content in $M_A$.
		Since this is the first time $D$ probes these cells, their contents remain the same as in $M_A$. 
		Bob then continues the simulation up to $i$-th query.
	\item (Stage $i$ - query answering)
		Bob simulates $D$ on query $Y_i$. 
		During the simulation, Bob \emph{pretends} that he has the right memory state for the query, even though he has skipped $I_A$, and only has received partial information about it.
		Then he outputs the same answer as $D$ does.
		Finally, Bob rolls back the memory to the version right before this query (after simulation described in Step 3). 
		That is, since the simulation on this query may be incorrect, Bob does not make any real changes to the memory in this step.
	\item
		As soon as Alice has sent $2\epsilon_v k\cdot w+1$ bits (where $w$ is the word-size), the players stop following the above steps, and output uniform random bit for all queries from this point.
\end{enumerate}

\paragraph*{Analyzing the Protocol}
It is easy to verify that Bob sends no message, and due to the last step, Alice always sends no more than $O(\epsilon_v k\cdot \log n)$ bits (word-size $w=\Theta(\log n)$).
Thus, $\cP_D$ has the first two properties claimed in the lemma statement.
In following, we are going to show that whenever $|P_A\cap P_B|\leq \epsilon_v k$ and $(1/2+\delta_v)$-fraction of queries in $I_B$ are correct, $\cP_D$ answers at least $(1/2+\delta_v-\epsilon_v)$-fraction of the queries correctly, which implies the third property.

In Step 2, Alice only sends Bob cells in $P_A\cap P_B$.
Moreover, each cell in the intersection will only be sent once - in the stage when it is probed by $D$ the first time.
Since sending the address and content of a cells takes $2w$ bits, as long as $|P_A\cap P_B|\leq \epsilon_v k$, the last step will not be triggered, and the players follow the first four steps.
Let us now focus on Step 4, query answering.
Although Bob pretends that he has the right memory state, which might not always hold, indeed for all queries during which $D$ does not probe any cell in $P_A$ that is not in Alice's messages, Bob \emph{will perform a correct simulation}.
That is, as long as $D$ does not probe any ``unknown'' cell in $P_A\cap P_B$, Bob will simulate $D$ correctly.
In the other words, each cell in $P_A\cap P_B$ can only lead to one incorrect query simulation among all $k$ queries.
When $|P_A\cap P_B|\leq \epsilon_v k$, on all but $\epsilon_v k$ queries, Bob's output agrees with the data structure.
Thus, at least $(1/2+\delta_v-\epsilon_v)$-fraction of the queries will be answered correctly, and this proves the lemma.
\end{proof}

To rule out the possibility of an efficient communication protocol for our problem, and prove Lemma~\ref{lem:oc_lower}, the main idea is to show that Bob has only learned very little information about the updates before each query $Y_i$.
Alice's message can only depend on $X$ and the previous queries, which are independent of $Y_i$.
Thus, the probability that Bob answers each query correctly must be close to $1/2$.
Finally, we obtain the desired probability bound from an application of the Azuma-Hoeffding inequality.

\begin{proof}[Proof of Lemma~\ref{lem:oc_lower}]
Let $R$ be the public random string, and $M_i$ be Alice's message in Stage $i$. 
Let $C_i$ be the indicator variable for correctly computing the $i$-th function $f_i(X, Y_i)$.
We first show that until Stage $i$, Bob has learned very little about $X$ even conditioned on $C_1,\ldots,C_{i-1}$, and thus could answer $Y_i$ correctly with probability barely greater than $1/2$. 
Formally, we will prove by induction on $i$ that
\[
	\Pr[C_i=1\mid C_1,\ldots,C_{i-1}]\leq \frac{1}{2}+O(\sqrt{\epsilon}+\sqrt{\epsilon_v}).
\]
Fix a sequence $c_1,\ldots,c_{i-1}\in\{0,1\}$.
For simplicity of notation, denote the event $C_1=c_1,\ldots,C_{i-1}=c_{i-1}$ by $W_c$.
By induction hypothesis, we have $\Pr[W_c]\geq 2^{-O(i)}\geq 2^{-O(k)}$.
Now conditioned on $W_c$, we upper bound the probability that $P$ correctly answers the $i$-th query:
\begin{align}
	&\Pr[P\textrm{ correctly computes }f_i(X, Y_i)\mid W_c] \notag \\
	\label{eqn:oc_lower1}
	&=\frac{1}{ns}\sum_{q=(l, b)\in [n]\times [s]}\Pr[P\textrm{ correctly computes }f_i(X, q)\mid W_c] \\
\intertext{Equality~\eqref{eqn:oc_lower1} is due to the fact that $Y_i$ is uniform and independent of the previous inputs.}
	\label{eqn:oc_lower2}
	&\leq \frac{1}{2}+\frac{1}{ns}\sum_{q=(l, b)\in [n]\times [s]}\E_{R,Y_1,\ldots,Y_{i-1},M_1,\ldots,M_i\mid W_c}\left|\Pr[f_i(X, q)=1\mid R,Y_1,\ldots,Y_{i-1},M_1,\ldots,M_i,W_c]-\frac{1}{2}\right| \\
\intertext{Inequality~\eqref{eqn:oc_lower2} holds because since Bob answers the query $q$
based only on $R,Y_1,\ldots,Y_{i-1},M_1,\ldots,M_i$, his advantage over
$\frac{1}{2}$ of answering correctly is at most the bias of the conditional
probability of $f_i(X, q)$.}
	\label{eqn:oc_lower3}
	&\leq\frac{1}{2}+\frac{1}{ns}\sum_{q=(l, b)\in [n]\times [s]}\Theta\left(\sqrt{1-H(f_i(X, q)\mid R,Y_1,\ldots,Y_{i-1},M_1,\ldots,M_i,W_c)}\right) \\
\intertext{Inequality~\eqref{eqn:oc_lower3} is due to Jensen's inequality and the fact
that for a binary random variable $Z$ such that
$\Pr[Z=1]=\frac{1}{2}\pm\epsilon$, its entropy is $H(Z)=1-\Theta(\epsilon^2)$.}
	\label{eqn:oc_lower4}
	&\leq \frac{1}{2}+\Theta\left(\sqrt{\frac{1}{ns}\sum_{q=(l, b)\in [n]\times [s]}(1-H(f_i(X, q)\mid R,Y_1,\ldots,Y_{i-1},M_1,\ldots,M_i,W_c)})\right).
\end{align}
Finally, Inequality~\eqref{eqn:oc_lower4} is from another application of Jensen's inequality.

Furthermore, we have
\begin{align}
	&\frac{1}{ns}\sum_{q=(l, b)\in [n]\times [s]}H(f_i(X, q)\mid R,Y_1,\ldots,Y_{i-1},M_1,\ldots,M_i,W_c) \notag \\
	\label{eqn:oc_lower5}
	&\geq \frac{1}{ns}\sum_{l\in [n]}H(a_{\leq l}(X, i)\mid R,Y_1,\ldots,Y_{i-1},M_1,\ldots,M_i,W_c) \\
	\label{eqn:oc_lower6}
	&\geq \frac{1}{ns}\sum_{o\in[n/k]}H(a_{\leq o}(X, i),a_{\leq o+n/k}(X, i), \ldots,a_{\leq o+(k-1)n/k}(X, i)\mid R,Y_1,\ldots,Y_{i-1},M_1,\ldots,M_i,W_c) \\
\intertext{$a_{\leq t}(X, i)$ is the product of first $t$ elements of $a$ right before $i$-th query of $I_B$ if the updates in $I_A$ is $X$, i.e., the group element that $q=(t, *)$ queries. Inequality~\eqref{eqn:oc_lower5} and \eqref{eqn:oc_lower6} is by the subadditivity of entropy and definition of the query function.}
	\label{eqn:oc_lower7}
	&\geq \frac{1}{ns}\sum_{o\in[n/k]}\left(H(X\mid R,Y_1,\ldots,Y_{i-1},M_1,\ldots,M_i,W_c)-s\right) \\
\intertext{Inequality~\eqref{eqn:oc_lower7} is by our construction of the update sequence.
The updates in $I_A$ are evenly spaced.
Thus, evenly spaced query can recover $X$ (possibly except one element, which has entropy at most $s$).}
	\label{eqn:oc_lower8}
	&=\frac{1}{ks}\cdot H(X\mid R,Y_1,\ldots,Y_{i-1},M_1,\ldots,M_i,W_c)-\frac{1}{k} \\
	\label{eqn:oc_lower9}
	&\geq \frac{1}{ks}\cdot \left(H(X\mid R,Y_1,\ldots,Y_{i-1},W_c)-H(M_1,\ldots,M_i\mid R,Y_1,\ldots,Y_{i-1},W_c)\right)-\frac{1}{k} \\
\intertext{Inequality~\eqref{eqn:oc_lower9} is by the chain-rule for conditional entropy.}
	\label{eqn:oc_lower10}
	&\geq \frac{1}{ks}\cdot H(X\mid R,Y_1,\ldots,Y_{i-1},W_c)-O(\epsilon_v)-\frac{1}{k} \\
\intertext{Inequality~\eqref{eqn:oc_lower10} is due to the fact that Alice sends no more than $O(\epsilon_v k\log n)$ bits and $s=\Theta(\log n)$.}
	\label{eqn:oc_lower11}
	&\geq \frac{1}{ks}\cdot \left(k\log |G|-\log\frac{1}{\Pr[W_c]}\right)-O(\epsilon_v)-\frac{1}{k} \\
\intertext{Inequality~\eqref{eqn:oc_lower11} is by the fact that $X$ is uniform and independent of $R,Y_1,\ldots,Y_{i-1}$. For uniform $X$, we have $H(X\mid W)\geq H(X)-\log\frac{1}{\Pr[W]}$ for any event $W$.}
	\label{eqn:oc_lower12}
	&\geq \frac{\log |G|}{s}-O\left(\frac{1}{s}\right)-O(\epsilon_v)-\frac{1}{k}\geq\frac{\log |G|}{s}-O(\epsilon_v+\epsilon), 
\end{align}
Inequality~\eqref{eqn:oc_lower12} is by the induction hypothesis that $\Pr[W_c]\geq 2^{-O(k)}$ and $\epsilon\geq \Omega(1/\log n)\gg 1/k$.

Combining the above inequalities, we have
\[
\begin{aligned}
	\Pr[P\textrm{ correctly computes }f_i(X, Y_i)\mid W_c]&\leq \frac{1}{2}+O\left(\sqrt{1-\frac{\log |G|}{s}+O(\epsilon_v+\epsilon)}\right) \\
	&\leq \frac{1}{2}+O\left(\sqrt{\epsilon}+\sqrt{\epsilon_v}\right).
\end{aligned}
\]

We have shown that conditioned on whether $P$ successfully computes first $i-1$ function values, the probability that it succeeds on the next is always upper bounded by $\frac{1}{2}+O\left(\sqrt{\epsilon}+\sqrt{\epsilon_v}\right)$.
Hence the random variables for the cumulative number of correct answers minus out cumulative upper bounds form a supermartingale, and we can apply the Azuma-Heoffding inequality~\cite{Hoeffding63}.
The probability that $(1/2+\delta_v-\epsilon_v)$-fraction of the function values are computed correctly is at most
\[
	\exp(-(\delta_v-O(\sqrt{\epsilon}+\sqrt{\epsilon_v}))^2\cdot k).
\]
This proves the lemma.
\end{proof}

\ifx\undefined\mainfile
\end{document}
\fi

\ifx\undefined\mainfile



\fi

\section{Dynamic Connectivity Lower Bound} \label{s:dynamic-connectivity}
In this section, we prove our lower bound for dynamic connectivity.

\begin{restate}[Theorem~\ref{thm:dynamic_graph_LB}]
There exists a distribution over $O(n)$ updates and queries for the dynamic connectivity problem, such that for any randomized cell-probe data structure $D$ with word-size $w=\Theta(\log n)$, which with probability $p$ answers at least a $(\frac{1}{2}+\delta)$-fraction of the queries correctly and spends $\epsilon n\log n$ total running time, we must have $p\leq \exp(-\delta^2n)$ as long as $\delta^2\gg 1/\log(1/\epsilon)$ and $\epsilon\geq \Omega(1/\log n)$ and $n$ is sufficiently large.
Moreover, the graph is always a forest throughout the sequence of updates.
\end{restate}

The high level strategy is very similar to the Group Range lower bound proof.
We first design a hard distribution (Section~\ref{sec_dcon_dist}).
Then we decompose the computation into many subproblems (Section~\ref{sec_dcon_step2}).
Finally, we prove via online communication for each subproblem, if the data structure is too efficient, then the probability of answering $(1/2+\delta)$-fraction of the queries correctly is exponentially small (the remaining subsections).

\subsection{Hard Distribution}
\label{sec_dcon_dist}

In this subsection, we describe the hard distribution $\cD$ for dynamic connectivity.
Without loss of generality, let us assume the number of vertices $n=B(B+1)+1$ for some integer $B$, and $B$ is a power of two. 
Our hard distribution is based on~\cite{PD06}, supported on sequences of $O(n)$ operations.
The data structure will have to maintain a graph on $n$ vertices: a special vertex $z$ and $B(B+1)$ vertices partitioned into $B+1$ layers $V_i=\{v_{i,j}: 1\leq j\leq B\}$ for $0\leq i\leq B$.
For any sequence of operations sampled from the hard distribution, the edges inserted will be either between the special vertex $z$ and some other vertex or between vertices in adjacent layers, $v_{i,j}\in V_i$ and $v_{i+1,j'}\in V_{i+1}$.
Moreover, the edges between any two adjacent layers will always form a perfect matching before every query.

Now let us describe the $O(n)$ random operations. 
We first initialize the graph by inserting $O(n)$ edges: Insert($v_{i-1,j}$, $v_{i,j}$) for all $1\leq i,j\leq B$.
That is, we first create a fixed graph as shown in Figure~\ref{fig:init}.
\begin{figure}
\centering
\begin{minipage}[t]{0.4\linewidth}
\centering
\begin{tikzpicture}
\foreach \i in {25,50,...,100}
	\foreach \j in {30,50,...,110}
		\node [draw,circle,inner sep=0pt,minimum size=3pt] at (\j pt,\i pt) (V\i\j) {};
\foreach \i in {25,50,...,100}
	\foreach \j/\k in {30/50,50/70,70/90,90/110}
		\draw (V\i\k) -- (V\i\j);
\node [draw,circle,inner sep=0pt,minimum size=3pt, label={$z$}] at (-15pt, 62.5pt) {};
\draw [rounded corners=5pt, thick, dashed] (23pt, 15pt) rectangle (37pt, 110pt);
\node at (30pt, 125pt) {$V_0$};
\end{tikzpicture}
\caption{The initialized graph when $B=4$.}\label{fig:init}
\end{minipage}
\hspace{30pt}
\begin{minipage}[t]{0.4\linewidth}
\begin{tikzpicture}
\foreach \i in {25,50,...,100}
  \foreach \j in {30,50,...,110}
    \node [draw,circle,inner sep=0pt,minimum size=3pt] at (\j pt,\i pt) (V\i\j) {};
\node [draw,circle,inner sep=0pt,minimum size=3pt, label={$z$}] at (-15pt, 62.5pt) (z) {};
\foreach \i in {25,50,75,100}
  \foreach \j/\k in {50/70,90/110}
    \draw (V\i\k) -- (V\i\j);
\foreach \i/\j in {25/50,50/100,75/25,100/75}
  \draw (V\i 30) -- (V\j 50);
\foreach \i/\j in {25/75,50/100,75/50,100/25}
  \draw (V\i 70) -- (V\j 90);

\draw (z) to[in=200,out=-5] (V7570);
\draw (z) to[in=-210,out=70] (V10070);
\end{tikzpicture}
\caption{A graph before querying whether a vertex is connected to either $v_{2,1}$ or $v_{2,2}$.}\label{fig:query}
\end{minipage}
\end{figure}
After the initialization, we start to update the graph by replacing the identity matchings between adjacent layers by random perfect matchings.
These $B$ matchings will be replaced in a fixed order, which we will specify below.
Due to technical reasons, the distribution of the new random matchings is deferred to Section~\ref{sec_dcom_comm_lower}.
The only property we will use for now is that the $B$ matchings are ``almost uniform and independent''.
More formally, we have the following proposition, whose proof is deferred to Section~\ref{sec_dcom_comm_lower} as well.
\begin{proposition}\label{prop_dist}
For $1\leq i\leq B$, denote by $M_i$ the random variable indicating the $i$-th updated matching in chronological order.
Then for any $0\leq k\leq B$, and any $m_1,\ldots,m_k$ in the support of the first $k$ matchings in $\cD$, conditioned on $M_i=m_i$ for all $1\leq i\leq k$, the distribution for the remaining $B-k$ matchings is a uniform distribution with support size at least $2^{-B-1}(B!)^{B-k}$.
\end{proposition}

The operations after the initialization are partitioned into $B$ \emph{operation blocks} $\cO_i$ for $0\leq i\leq B-1$.
In operation block $\cO_i$, we will focus on the layer $V_{\rev{\log B}{i}}$.\footnote{Write $i$ as a $t$-bit binary number, $\rev{t}{i}$ is the number with its bits reversed. See Section~\ref{s:group-range} for the definition.}
For simplicity of notation, let $r_i=\rev{\log B}{i}$.
We first replace the matching between $V_{r_i}$ and $V_{r_i+1}$.
That is, we first $\del(v_{r_i,j},v_{r_i+1,j})$ for all $1\leq j\leq B$.
Then, we generate a random permutation $F: [B]\rightarrow [B]$, and $\ins(v_{r_i,j},v_{r_i+1,F(j)})$.
These permutations are sampled from a distribution satisfying the property in Proposition~\ref{prop_dist}.
After replacing the matching, we do $B$ queries, each of which is of form ``whether vertex $u$ is connected to any vertex in the set $\{v_{r_i,j}: 1\leq j\leq B/2\}$ (the top half of $V_{r_i}$).''
They are not standard connectivity queries.
However, it is possible to implement them with a few extra insertions and deletions.
To do this, we begin by inserting $B/2$ edges: $\ins(z, v_{r_i,j})$ for all $1\leq j\leq B/2$.
Next, we do $B$ queries $\con(z, u)$ for independent and uniformly random $u$.
Finally, we delete the $B/2$ edges inserted earlier: $\del(z, v_{r_i,j})$ for all $1\leq j\leq B/2$ (see Figure~\ref{fig:query}).
Each $\cO_i$ consists of $4B$ operations.
Thus, the sequence has $O(n)$ operations in total.

\subsection{Identifying Key Subproblems}
\label{sec_dcon_step2}

In this section, we relate the lower bound on subproblems that we will prove with our desired data structure lower bound, Theorem~\ref{thm:dynamic_graph_LB}. As a reminder, here is the reduction theorem that we will be using:
\begin{restate}[Theorem~\ref{thm:step2}]
  Suppose that there is a data structure problem along with a hard distribution for it over sequences of $n$ blocks consisting of $n_b$ operations each, for a total of $N = n n_b$ operations. Next, suppose there exists a constant $c \in (0, 1)$, value $\epsilon_0>0$, and a bivariate convex function $g(x, y)$, whose value is non-decreasing in $x$ and non-increasing in $y$, so that the following is true: for any data structure $D$, any subproblem $(I_A(v), I_B(v))$ where $I_B(v)$ consists of $k \ge n^{1-c}$ blocks, any $\epsilon_v \geq 0$ and $\delta_v\in [0,1/2]$, the probability \emph{conditioned on} all operations before $I_A(v)$ that the following hold:
  \begin{itemize}
    \item $\SharedProbes \le \epsilon_v \cdot k n_b$,
    \item $D$ answers a $(\frac12 + \delta_v)$-fraction of queries in $I_B(v)$ correctly,
  \end{itemize}
  is at most $\exp \left( -g(\delta_v,\epsilon_v) k n_b \right)$.
  Then the probability that all the following hold:
  \begin{itemize}
    \item $D$ probes at most $\epsilon N \log n$ cells,
    \item $D$ answers a $(\frac12 + \delta)$-fraction of all queries correctly,
  \end{itemize}
  is at most $\exp(n^c\cdot \log N) \cdot \exp\left(-g(\delta-3/\sqrt{c\log n}, \epsilon/c) \cdot N\right)$ as long as $\delta\geq 3/\sqrt{c\log n}$.
\end{restate}

We will be proving the following bound on subproblems in the next subsection:
\begin{lemma}\label{lem_dcon_int}
Suppose we have two intervals $I_A(v)$ and $I_B(v)$ consisting of $k$ operation blocks each.
Then the probability conditioned on all operations $O$ before $I_A(v)$ that
\begin{itemize}
	\item $|P_A(v)\cap P_B(v)|\leq \epsilon_v\cdot kB$ and
	\item $(\frac12+\delta_v)$-fraction of the queries in $I_B(v)$ are answered correctly
\end{itemize}
is at most $\exp((-\delta_v^2+\beta/\log (1/ \epsilon_v))kB)$ for some constant $\beta>0$, as long as $k\geq B^{1/8}$, $\frac{3}{w}<\epsilon_v<\frac{\log n}{3w}$ and $\epsilon_v<\delta_v$, where $w$ is the word-size.
\end{lemma}

Since Theorem~\ref{thm:step2} has a convexity requirement, we will also need the following technical lemma about the convexity of our error function:
\begin{lemma}\label{lem_dcon_convex}
  For any $\beta > 0$, the function $g_0(x, y) = \max \{0, x^2 - \frac{\beta}{\ln 1/y} - \frac{\beta}{\ln w/3} \}$ is convex over $(x, y) \in [0, 1] \times (0, 1/e^2]$.
\end{lemma}

\begin{proof}
  The max of two convex functions is convex. $g_1(x, y) = 0$ is a constant function, so it is convex. It suffices to prove that $g_2 = x^2 - \frac{\beta}{\ln 1/y}$ is convex as well. We do so by showing that it is the sum of two convex functions: $g_3(x, y) = x^2-\frac{\beta}{\ln w/3}$ and $g_4(x, y) =  -\frac{\beta}{\ln 1/y}$. It is easy to see that $g_3$ is convex; it is really a single-variable function with second derivative $g_1''(x) = 2$. Hence our main task is to show $g_4$ is convex as well.
  
  We first compute the second derivative of $g_4$, since it also only depends on a single variable. We use a combination of chain, product, and quotient rules:
  \begin{align*}
    g_4(y) &= - \frac{\beta}{\ln 1/y} \\
    g_4'(y) &= - \frac{0 - \beta (y) (-1/y^2) }{ \ln^2 1/y } \\
      &= - \frac{\beta}{ y \ln^2 1/y } \\
    g_4''(y) &= - \frac{0 - \beta \left[ (y) (2 \ln 1/y) (y) (-1/y^2) + (1) (\ln^2 1/y) \right]}{y^2 \ln^4 1/y} \\
      &= \frac{\beta \left[ -2 \ln 1/y + \ln^2 1/y \right]}{y^2 \ln^4 1/y}
  \end{align*}
  Note that when $y \in (0, 1)$, the final denominator is always positive. Additionally, since $\beta > 0$, we only care about whether:
  \begin{align*}
    -2 \ln 1/y + \ln^2 1/y &\ge 0 \\
    -2 + \ln 1/y &\ge 0 \\
    \ln 1/y &\ge 2 \\
    1/y &\ge e^2 \\
    y &\le 1/e^2
  \end{align*}
  Hence for $y \in (0, 1/e^2]$, $g_4$ is convex. Combining our convexity claims, the original function $g_0$ is convex over the desired range.
\end{proof}

Now, we are ready to show that we meet the condition for Theorem~\ref{thm:step2} and prove Theorem~\ref{thm:dynamic_graph_LB}.
\begin{proof}[Proof of Theorem~\ref{thm:dynamic_graph_LB}]
 Note that the requirement of Lemma~\ref{lem_dcon_int} that $k \ge B^{1/8}$ can be satifsfied by choosing $c = 7/8$ and $n$, the number of blocks of operations, to be $B$.
We define
\[
	g(x, y) := \begin{cases}\max\{0,x^2- \frac{\beta}{\ln w/3}\} & y=0 \\ \max\{0, x^2 - \frac{\beta}{\ln 1/y} - \frac{\beta}{\ln w/3}\} & y\in (0,1/e^2) \\ 0 & y \geq 1/e^2\end{cases}
\]
for some large enough constant $\beta$.
It is not hard to verify that $g$ is continuous, non-decreasing in $x$ and non-increasing in $y$.
By Lemma~\ref{lem_dcon_convex}, $g$ is also convex over $[0,1]\times [0,\infty)$.

Furthermore, we claim that the probability conditioned on all operations $O$ before $I_A(v)$ that
\begin{itemize}
	\item $|P_A(v)\cap P_B(v)|\leq \epsilon_v\cdot kB$ and
	\item $(\frac12+\delta_v)$-fraction of the queries in $I_B(v)$ are answered correctly
\end{itemize}
is at most $\exp(-g(\delta_v,\epsilon_v)kB)$.

Since $w = \Theta (\log n)$, we make $\beta$ large enough so that when $\epsilon_v\geq \frac{\log n}{3w}$, $g(\delta_v,\epsilon_v)=0$.
Thus, the claim is trivially true in this case.
When $\epsilon_v\geq \delta_v$, since $\beta$ is large enough, we have
\[
	\delta_v^2-\beta/\log (1/\epsilon_v)\leq \delta_v^2-\beta/\log (1/\delta_v)<0.
\]
That is, $g(\delta_v,\epsilon_v)=0$ and the claim is true.
When $\epsilon_v<\delta_v$ and $\frac{3}{w}<\epsilon_v<\frac{\log n}{3w}$, 
$g(\delta_v,\epsilon_v)\leq \max\{0,\delta_v^2-\beta/\log(1/\epsilon_v)\}$,
the claim is true by Lemma~\ref{lem_dcon_int}.
Finally, when $\epsilon_v\leq 3/w$, $g(\delta_v,\epsilon_v)\leq \max\{0,\delta_v^2-\beta/\log(w/3)\}$, and the claim is true by monotonicity and Lemma~\ref{lem_dcon_int}.




By Theorem~\ref{thm:step2}, $D$ succeeds with probability at most:
\[
  \exp(B^{7/8} \cdot \log n) \cdot \exp\left(- \left( \left(\delta-\frac{3}{\sqrt{\frac78 \log n}} \right)^2 - \frac{\beta}{\log \frac{7}{16\epsilon}} \right) \cdot n\right).
\]

Since $\delta^2 \gg 1 / \log (1 / \epsilon)$, $\epsilon\geq \Omega(1/\log n)$, and $n$ sufficiently large, this probability is upper bounded by $\exp(-\delta^2 n)$.
This proves the theorem.
\end{proof}

\subsection{Communication Game}\label{sec_dcon_game}

In this subsection, we will prove Lemma~\ref{lem_dcon_int} using online communication.

\paragraph{Communication Game.} We define an online communication game for each interval pair $(I_A(v), I_B(v))$.
Let $O$ be the operations before $I_A$, $Q_A$ be the queries in $I_A$ and $U_B$ be the updates in $I_B$.
In the online communication game $G=G(v, O, Q_A, U_B)$, Alice's input $X$ is the updates in $I_A$, Bob's input $Y_i$ is the $i^{th}$ query in $I_B$.
The goal of stage $i$ is to answer the $i^{th}$ query.

The inputs $(X, Y_1,\ldots)$ are sampled according to the distribution $\cD$ for the data structure and the known operations, i.e. they are jointly sampled conditioned on $O$, $Q_A$ and $U_B$.
It is easy to verify that all the queries $Y_i$ are independent, and they are independent of $X$.

To prove Lemma~\ref{lem_dcon_int}, we will apply a similar strategy as the proof for the Group Range lower bound.
We will first apply Lemma~\ref{lem:oc_upper} to obtain an efficient communication protocol from an efficient data structure, which roughly preserves the fraction of correct queries.
Then to prove the communication lower bound, we are going to apply the following generalized Chernoff bound:

\begin{theorem}[\cite{PS97,IK10}]\label{thm:gen_chernoff}
Let $X_1,\ldots,X_n$ be $n$ Boolean random variables.
Suppose that there are $0\leq \mu_i\leq 1$, for $1\leq i\leq n$, and $\lambda>0$, for all $S\subseteq [n]$,
\[
	\Pr\left[\wedge_{i\in S} X_i=1\right]\leq \lambda\cdot \prod_{i\in S}\mu_i.
\]
Let $\mu=(1/n)\sum_{i=1}^n \mu_i$.
Then for any $1\geq\delta\geq\mu$, 
\[
	\Pr\left[\sum_{i=1}^n X_i\geq \delta n\right]\leq \lambda\cdot 2^{-n D(\delta||\mu)},
\]
where $D(\delta||\mu)=\delta\log\frac{\delta}{\mu}+(1-\delta)\log\frac{1-\delta}{1-\mu}$ is the binary relative entropy function.
\end{theorem}
\begin{remark}\label{rem_chernoff}
It is easy to verify that $D((1+\alpha)\mu||\mu)\geq \frac{1}{2}\alpha\log (1+\alpha)\cdot \mu$.
Thus, we also have
\[
	\Pr\left[\sum_{i=1}^n X_i\geq (1+\alpha)\mu n\right]\leq \lambda\cdot 2^{-\frac{1}{2}\alpha\log(1+\alpha) \mu n}.
\]
\end{remark}

That is, in order to upper bound the probability the protocol answers $(1/2+\delta)$-fraction of the queries correctly, it suffices to show that for every subset $S$ of the queries, the probability that they are all correct is very close to $2^{-|S|}$.
In fact, it even suffices to prove it when $S$ is the set of all queries in $I_B$.

\begin{lemma}\label{lem_dcon_all}
Let $G(v,O,Q_A,U_B)$ be a communication game defined as above with $k$ operation blocks in $I_A(v)$ and $I_B(v)$.
If $k\geq B^{1/8}$ and the min-entropy of input $X$ is at least $k\log B!-2kB$,\footnote{The min-entropy of $X$ is at least $c$ if and only if no singleton $x$ has $\Pr[X=x]\geq 2^{-c}$.} then
for any protocol $P$ and $\epsilon_v$ where:
\begin{enumerate}
\item Alice sends $\epsilon_v kB\cdot \log B$ bits, and
\item Bob sends no message
\end{enumerate}
must have
\[
	\Pr[P\textrm{ answers \emph{all} queries correctly}] \leq 2^{-(1-\gamma) kB},
\]
where $\gamma=24/\log (1/\epsilon_v)$, as long as $\frac{9}{\log B}<\epsilon_v<\frac{1}{2}$.
\end{lemma}

Assuming the above lemma (which we will prove in Section~\ref{sec_dcom_comm_lower}), we will be able to prove the probability of answering $(1/2+\alpha)$-fraction of the queries correctly is tiny.

\begin{lemma}\label{lem_dcon_comm_lower}
Let $G(v,O,Q_A,U_B)$ be a communication game defined as above with $k$ operation blocks in $I_A(v)$ and $I_B(v)$.
If $k\geq B^{1/8}$ and the min-entropy of input $X$ is at least $k\log B!-2kB$, then
for any protocol $P$, $\delta_v$ and $\epsilon_v$ where:
\begin{enumerate}
\item Alice sends $\epsilon_v kB\cdot \log B$ bits, and
\item Bob sends no message
\end{enumerate}
must have
\[
	\Pr[P\textrm{ answers $(\frac12+\alpha)$-fraction queries correctly}] \leq 2^{\left(-2\alpha^2+24/\log (1/\epsilon_v)\right)kB},
\]
as long as $\frac{9}{\log B}<\epsilon_v<\frac{1}{2}$ and $\epsilon_v<\delta_v$.
\end{lemma}

\begin{proof}[Proof of Lemma~\ref{lem_dcon_comm_lower}]
For $1\leq i\leq kB$, let $X_i$ be the indicator variable for the event that the $i$-th query in $I_B$ is answered correctly by protocol $P$.
For any $S\subseteq [kB]$, define $P_S$ to be the protocol such that $P_S$ does exactly the same thing as $P$ except that it outputs independent random bits on every query that is not in $S$.
Then the probability that $P_S$ answers all queries correctly is exactly $\Pr[\wedge_{i\in S}X_i=1]\cdot 2^{|S|-kB}$.
On the other hand, by Lemma~\ref{lem_dcon_all}, we have this probability is at most $2^{-(1-\gamma)kB}$ for $\gamma=24/\log (1/\epsilon_v)$.
Therefore, 
\[
	\Pr[\wedge_{i\in S}X_i=1]\leq 2^{\gamma kB}\cdot 2^{-|S|}.
\]

By Theorem~\ref{thm:gen_chernoff}, we have
\[
\begin{aligned}
	\Pr\left[\sum_{i=1}^{kB} X_i\geq (\frac12+\alpha)kB\right]&\leq 2^{\gamma kB}\cdot 2^{-D(\frac{1}{2}+\alpha||\frac12)kB} \\
	&\leq 2^{\left(-2\alpha^2+24/\log (1/\epsilon_v)\right)kB}
\end{aligned}
\]
by setting $\mu_i=1/2$, $\delta=\frac12+\alpha$ and $\lambda=2^{\gamma kB}$.
\end{proof}

\begin{proof}[Proof of Lemma~\ref{lem_dcon_int}]
Fix one sequence $O$ of operations before $I_A$ in the support of $\cD$.
By Proposition~\ref{prop_dist}, conditioned on $O$, the sequence of remaining updates ($X$, $U_B$ and the updates after the intervals $R$) has support size at least $2^{-B-1}(B!)^{B-|O|}$.
Then if we sample a random $U_B$, the probability that the remaining updates have a small support is small:
\[
\begin{aligned}
	&\sum_{U_B:|\mathrm{supp}(X, R\mid O,U_B)|<2^{-2kB}\cdot (B!)^{|X|+|R|}}\Pr[U_B\mid O] \\
	&<\sum_{U_B:|\mathrm{supp}(X, R\mid O,U_B)|<2^{-2kB}\cdot (B!)^{|X|+|R|}} \frac{2^{-2kB}\cdot (B!)^{|X|+|R|}}{2^{-B-1}(B!)^{B-|O|}}\\
	&\leq 2^{-2kB+B+1},
\end{aligned}
\]
where $\mathrm{supp}(X, R\mid O, U_B)$ is the support of $X$ and $R$ conditioned on $O$ and $U_B$.
On the other hand, for every possible assignment $x$ to $X$, we have
\[
\begin{aligned}
	\Pr[X=x\mid O, U_B]&=\sum_r \Pr[X=x, R=r\mid O, U_B] \\
	&\leq (B!)^{|R|}/|\mathrm{supp}(X, R\mid O,U_B)|.
\end{aligned}
\]
When $|\mathrm{supp}(X, R\mid O,U_B)|\geq 2^{-2kB}\cdot (B!)^{|X|+|R|}$, we have
\[
	\Pr[X=x\mid O, U_B]\leq 2^{2kB}\cdot (B!)^{-|X|}.
\]
That is, the min-entropy of $X$ conditioned on $O$ and $U_B$ is at least $k\log B!-2kB$.
Thus, over the randomness of $U_B$, the min-entropy of $X$ is at least $k\log B!-2kB$ with probability at least $1-2^{-2kB+B+1}$.

Finally by Lemma~\ref{lem:oc_upper}, Lemma~\ref{lem_dcon_comm_lower} and union bound, the probability conditioned on $O$ that
\begin{itemize}
	\item $|P_A(v)\cap P_B(v)|\leq \epsilon_v\cdot k$ and
	\item all queries in $I_B(v)$ are answered correctly
\end{itemize}
is at most $$2^{-2kB+B+1}+2^{\left(-2(\delta_v-\epsilon_v)^2+24/\log (\log n/3w \epsilon_v)\right)kB}<\exp((-\delta_v^2+\beta/\log (1/\epsilon_v))kB)$$ for some constant $\beta>0$.
\end{proof}

\subsection{The Monologue Lemma}

\begin{figure}
\centering
\begin{tikzpicture}[%
  scale=0.45,
  auto,
  vertex/.style={
    circle,
    draw=blue,
    thick,
    fill=blue!20,
    align=center,
  }
  ]
  \node[vertex] (v) at (0, 0) {};
  \node[vertex] (v0) at (-12, -4) {};
  \node[vertex] (v1) at (-4, -4) {};
  \node[vertex] (v2) at (4, -4) {};
  \node[vertex] (v3) at (12, -4) {};
  \node[vertex] (v00) at (-15, -8) {};
  \node[vertex] (v01) at (-13, -8) {};
  \node[vertex] (v02) at (-11, -8) {};
  \node[vertex] (v03) at (-9, -8) {};
  \node[vertex] (v10) at (-7, -8) {};
  \node[vertex] (v11) at (-5, -8) {};
  \node[vertex] (v12) at (-3, -8) {};
  \node[vertex] (v13) at (-1, -8) {};
  \node[vertex] (v20) at (1, -8) {};
  \node[vertex] (v21) at (3, -8) {};
  \node[vertex] (v22) at (5, -8) {};
  \node[vertex] (v23) at (7, -8) {};
  \node[vertex] (v30) at (9, -8) {};
  \node[vertex] (v31) at (11, -8) {};
  \node[vertex] (v32) at (13, -8) {};
  \node[vertex] (v33) at (15, -8) {};
  
  \draw[blue, solid] (v) -- node [black, below right] {$q_1$} (v0);
  \draw[red, dashed] (v) -- node [black, below right] {$q_2$} (v1);
  \draw[blue, solid] (v) -- node [black, below left] {$q_3$} (v2);
  \draw[red, dashed] (v) -- node [black, below left] {$q_4$} (v3);
  
  \draw[red, dashed] (v0) -- (v00);
  \draw[red, dashed] (v0) -- (v01);
  \draw[blue, solid] (v0) -- (v02);
  \draw[red, dashed] (v0) -- (v03);
  \draw[blue, solid] (v1) -- (v10);
  \draw[red, dashed] (v1) -- (v11);
  \draw[blue, solid] (v1) -- (v12);
  \draw[blue, solid] (v1) -- (v13);
  \draw[red, dashed] (v2) -- (v20);
  \draw[red, dashed] (v2) -- (v21);
  \draw[blue, solid] (v2) -- (v22);
  \draw[blue, solid] (v2) -- (v23);
  \draw[blue, solid] (v3) -- (v30);
  \draw[blue, solid] (v3) -- (v31);
  \draw[blue, solid] (v3) -- (v32);
  \draw[red, dashed] (v3) -- (v33);
  
  \node (ell0) at (-17, 0) {};
  \node (ell1) at (-17, -4) {};
  \node (ell2) at (-17, -8) {};
  \draw[black, thick, solid] ($(ell0) + (-1,0)$) -- ($(ell0) + (+1, 0)$);
  \draw[black, thick, solid] ($(ell1) + (-1,0)$) -- ($(ell1) + (+1, 0)$);
  \draw[black, thick, solid] ($(ell2) + (-1,0)$) -- ($(ell2) + (+1, 0)$);
  \draw[black, thick, solid] (ell0.center) -- (ell1.center) node[midway, left] {$Y_1$};
  \draw[black, thick, solid] (ell1.center) -- (ell2.center) node[midway, left] {$Y_2$};
\end{tikzpicture}
\caption{After Alice sees her input $X$, she can simulate protocol $P$ for all
  possible sequences $Y_1, \ldots, Y_k$ that Bob could receive. She can organize
  these results into a tree, and she knows whether Bob will answer a particular
  input $Y_i = q_j$ correctly (solid blue line) or incorrectly (dashed red
  line). In the tree above, all of Bob's queries are answered correctly only
  when his input is one of three specific sequences.}
\label{f:bob-tree}
\end{figure}
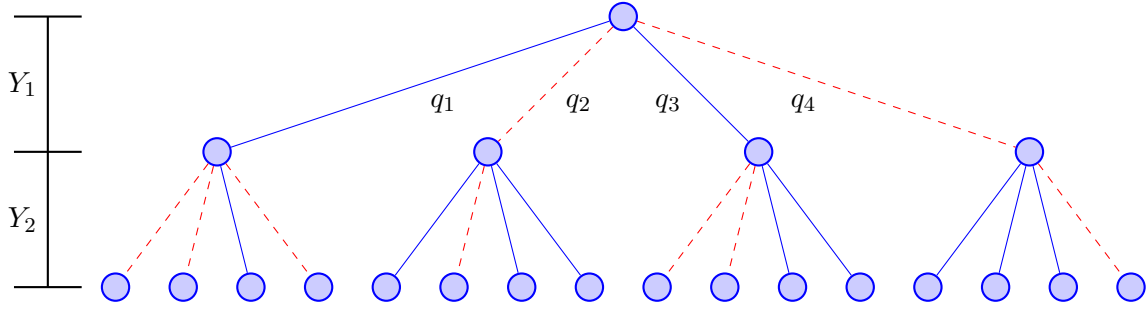

One of the running themes in this work is that online protocols are much easier
to reason about if we only need to reason about a single round. Keeping with
this trend, we now prove that if an online protocol only has Alice speak (i.e.
she is monologuing), then she might as well do it at the very beginning. 
\begin{lemma}[Monologue Lemma]\label{lem_mono}
  Suppose we have a problem in our online communication model and associated
  input distribution $\mathcal{D}$ over $\cX \times \cY^k$ with the property
  that Alice's input $X$ and Bob's inputs $Y_1, \ldots, Y_k$ are independent.
  Furthermore, suppose we have a randomized protocol $P$ such that for all of
  Alice's inputs $X \in \cX$, with at least probability $p$ (over Bob's inputs)
  all of the following events occur (the protocol ``succeeds''):
  \begin{enumerate}
    \item Only Alice talks.
    \item At most $C$ bits are sent.
    \item All of Bob's queries are answered correctly.
  \end{enumerate}
  Then there is another randomized protocol $P'$ with the following properties:
  \begin{enumerate}
    \item Only Alice talks, and she only does so in the first stage.
    \item $C + \log 1/p + O(\log\log 1/p)$ bits are sent in
      expectation.
    \item All of Bob's queries are answered correctly with probability at least
      $p$.
  \end{enumerate}
\end{lemma}

\begin{proof}
  We first assume without loss of generality that $P$ is deterministic; we have
  an input distribution and can apply Yao's minimax principle, even if $P$ is a
  public-coin protocol.

  Here is the protocol $P'$ on input $X, Y_1, \ldots, Y_k$:
  \begin{enumerate}
    \item Alice and Bob look at the public randomness and interpret it as a
      sequence of fake draws of Bob's input from the input distribution
      $\mathcal{D}$. Alice computes the first draw which, when combined with her
      input $X$, would be succeeded on by $P$.
    \item In this fake draw, Bob would get some inputs $\tilde{Y}_1, \ldots,
      \tilde{Y}_k$. Alice knows these inputs, so she can simulate protocol $P$
      on her real input combined with the fake draw
      $(X, \tilde{Y}_1, \ldots, \tilde{Y}_k)$. She sends the index of the fake
      draw and the resulting transcript to Bob as her only communication for
      $P'$.
    \item In round $i$, Bob receives his real input $Y_i$. Bob uses his
      knowledge of the fake input and resulting transcript to simulate protocol
      $P$ on the hybrid input $(X, \tilde{Y}_1, \ldots, \tilde{Y}_{i-1}, Y_i)$
      up to round $i$. He outputs whatever it does for round $i$.
  \end{enumerate}

  By construction, only Alice talks and she only does so in the first round.
  Let us analyze Alice's message length. Since the chance of finding a
  satisfactory fake draw is at least $p$, the expected index is at most
  $\frac{1}{p}$. Note that the index can be unbounded. A standard approach to
  encoded an unbounded number $x$ takes $\log x+O(\log\log x)$ bits. By
  Jensen's inequality and the concavity of the log function, the expected
  number of bits to send an index is $\log 1/p+O(\log\log 1/p)$. Since we find
  an input that protocol $P$ succeeds on, the transcript component is at most
  $C$ bits.

  The difficult part of the proof is proving the last property. We want to show
  that we lose nothing by performing this hybrid input procedure (and it is here
  that we will use the fact that Bob's inputs are independent). We will show
  that the probability that $P'$ succeeds on all queries is at least the
  probability that $P$ succeeds on all queries. This would show that $P'$
  answers all queries correctly with probability at least $p$, completing the
  proof.
  
  Fix Alice's input $X \in \cX$. We think of Bob's possible input sequences
  as a complete $k+1$-layer, $|\cY|$-ary tree. This tree is depicted in
  Figure~\ref{f:bob-tree}. Every root-to-leaf path represents a possible input
  sequence; the edge that the path takes from layer $i$ to $i+1$ corresponds to
  $Y_i$. Denote by $P(i, Y_i)$ the probability that Bob's $i^{th}$ input is
  $Y_i$. We can write a root-to-leaf as an input sequence for Bob:
  $L = (Y_1, \ldots, Y_k)$. The probability that $L$ occurs is:
  \[
    \Pr[L] = \prod_{i=1}^k P(i, Y_i).
  \]
  Suppose $v$ is a node in this tree. We denote the probability that Bob answers
  his next query correctly conditioned on being at $v$ by $C(v)$. The
  probability that our protocol succeeds when it uses the fake input
  $L = (\tilde{Y}_1, \ldots, \tilde{Y}_k)$ is exactly:
  \[
    \prod_{v \in L} C(v).
  \]
  The probability that our protocol is correct when it uses a \emph{random} fake
  input $L$ is:
  \[
    \frac{\sum_{L\textrm{ succeeds}} \left[ \Pr[L]\cdot \prod_{v \in L} C(v) \right]}{\sum_{L\textrm{ succeeds}}\Pr[L]}
  \]
  We want to show this is at least the success probability of the original protocol, which is:
  \[
    \sum_{L\textrm{ succeeds}} \Pr[L].
  \]
  That is, we want to show
  \[
    \sum_{L\textrm{ succeeds}}\Pr[L]\cdot \prod_{v \in L} C(v)\geq \left(\sum_{L\textrm{ succeeds}}\Pr[L]\right)^2
  \]
  Our plan is to prove this by inducting over the tree. For any node $u$ in the
  tree, define:
  \begin{align*}
    g(u)&:=\sum_{L\textrm{ succeeds, descendant of $u$}} \left[ \Pr[L\mid u]\cdot \prod_{v \in L, \textrm{ descendant of }u} C(v) \right] \\
    P(u)&:=\sum_{L\textrm{ succeeds, descendant of $u$}} \Pr[L\mid u],
  \end{align*}
  where $\Pr[L\mid u]$ is the probability that $L$ occurs conditioned on reaching node $u$.
  For $L=(Y_1,\ldots,Y_k)$ and $u$ depth $d$, we have
  \[
  	\Pr[L\mid u]=\prod_{i=d}^k P(i, Y_i).
  \]

  Our inductive hypothesis is that $g(u) \geq P(u)^2$. When $u$ is the root of
  the tree, we get the original claim.
  
  The base case is easy: if $u$ is a leaf, then both sides are $0$ or $1$,
  depending on whether $u$'s root-to-leaf path succeeds or not. Now, suppose
  the hypothesis holds for all children of $u$, and we want to show it holds for
  $u$ in layer $i$. Let the children of $u$ on at least one successful root-to-leaf path be
  $u_1, u_2, \ldots, u_m$ which correspond to Bob's $i^{th}$ input being
  $q_1, q_2, \ldots, q_m$, respectively. We finish with an application of
  Cauchy-Schwarz:
  \begin{align*}
    g(u) &=\sum_{j=1}^m P(i, q_j) \cdot C(u) \cdot g(u_i) \\
      &\ge \sum_{j=1}^m P(i, q_j) \cdot C(u) \cdot P(u_i)^2 \\
      &= C(u) \cdot \left(\sum_{j=1}^m P(i, q_j) \cdot P(u_i)^2\right) \\
      &\ge \left(\sum_{j=1}^m P(i, q_j) \right) \left(\sum_{j=1}^m P(i, q_j) \cdot P(u_i)^2\right)\\
      &\ge \left(\sum_{j=1}^m P(i, q_j) \cdot P(u_i)\right)^2 \\
      &= P(u)^2.
  \end{align*}
  Hence we have proved our inductive hypothesis for all nodes in the tree and in
  particular, the root. This proves that protocol $P'$ has at least a $p$
  probability overall of being correct on all queries. Hence we managed to prove
  all the desired properties about $P'$.
\end{proof}

Our corollary removes the assumption that for every one of Alice's inputs, the
protocol $P$ must maintain the probability of success, instead it only
requires the protocol to have a good overall success probability. We tack on
the restriction that the functions $f_1, \ldots, f_k$ are binary, but it is
possible to prove a version for larger domains.

\begin{corollary}[Monologue Corollary]\label{cor_mono}
  Suppose we have a problem in our online communication model and associated
  input distribution $\mathcal{D}$ over $\cX \times \cY^k$ with the property
  that Alice's input $X$ and Bob's inputs $Y_1, \ldots, Y_k$ are independent.
  Furthermore, suppose we have a randomized protocol $P$ with the following
  properties:
  \begin{enumerate}
    \item Only Alice talks.
    \item At most $C$ bits are sent.
  \end{enumerate}
  If all functions $f_1, \ldots, f_k$ are boolean-valued, then there is another randomized protocol $P'$ with the following properties:
  \begin{enumerate}
    \item Only Alice talks, and she only does so in the first round.
    \item $C + k + O(\log k)$ bits are sent in expectation.
    \item The probability that $P'$ answers all queries correctly is at least much as that of $P$.
  \end{enumerate}
\end{corollary}

\begin{proof}
To apply the Monologue lemma, we must have a lower bound on the success
probability for every input $X$. However for any protocol, we can always have
Alice use one bit in the very beginning to indicate whether, for her input
$X$, the protocol has at least a $2^{-k}$ chance of getting all queries
correct (over the randomness of Bob's inputs). If it does have such a chance,
then they proceed normally. Otherwise, Alice says no more and Bob outputs
uniformly random bits for every query; then he has at least a $2^{-k}$
probability since the functions are boolean-valued.  This increases the number
of bits transmitted by one, while now for every input the success probability
is at least $2^{-k}$. The corollary then follows by the Monologue lemma.
\end{proof}

\subsection{Communication Lower Bound}\label{sec_dcom_comm_lower}

In this subsection, we prove Lemma~\ref{lem_dcon_all}.
The dynamic graph has $B$ matchings between $B+1$ layers of size $B$ each.
$I_A$ and $I_B$ both consist of $k$ operation blocks and start at $c k$-th and $(c+1) k$-th operation blocks in the sequence respectively for some integer $c$.
By the property of $\mathrm{rev}$ and the fact that $k$ is a power of two, Alice's and Bob's inputs have the following properties:
\begin{itemize}
	\item $k$ matchings are updated in $I_A$, which we will refer to as the \emph{missing matchings}, as they are the only information about the graph that Bob does not know;
	\item $k$ matchings are updated in $I_B$, which we will refer to as the \emph{replaced matchings};
	\item after each matching update in $I_B$, we ``query'' the layer adjacent to it, i.e., $B$ random queries of form whether a node is connected to the top half of this layer are asked; we refer to these layers as the \emph{queried layers};
	\item the $k$ missing matchings and the $k$ replaced matchings are interleaved.
\end{itemize}

To prove Lemma~\ref{lem_dcon_all}, the first step is to apply Corollary~\ref{cor_mono}.
We may assume without loss of generality, Alice only speaks before the first stage and her message length is at most $\epsilon_vkB\cdot\log B+\log 1/2^{-kB}+1\leq (1+\epsilon_v\log B)kB+O(\log B)$.

The high-level idea to prove Lemma~\ref{lem_dcon_all} is to let Alice send the first message about her input (i.e., the missing matchings) of length at most $(1+\epsilon_v\log B)kB+O(\log B)$.
Since Bob is able to complete the protocol with no further communication, a random sequence of $kB$ queries can be answered correctly based solely on this message with probability $2^{-(1-\gamma) kB}$.
The players then treat the public random string as repeated samples of queries.
On average, there is one all-correct sample of queries in every $2^{(1-\gamma) kb}$ samples.
Thus, it takes only about $(1-\gamma) kB$ bits for Alice to specify each sample that would be answered all correctly by Bob.
Ideally, every $kB$ queries may reveal $kB$ bits of information about Alice's input.
That is, in the ideal situation, Alice will be able to save about $\gamma kB$ bits each time at the cost of sending $(1+\epsilon_v\log B)kB$ extra bits in the beginning.
If Alice managed to repeat it much more than $(1+\epsilon_v\log B)/\gamma$ times, and each time reveal about $kB$ extra bits of information, she would have compressed her input below the information theoretical lower bound, yielding a contradiction.

Of course, not every set of $x$ queries reveals about $x$ bits of information.
In the following, we derive a sufficient condition for this property (Lemma~\ref{lem_enc_perm}), and show that this condition happens with extremely high probability (Lemma~\ref{lem_regular}, Lemma~\ref{lem_even_spd}).

By the construction of our hard distribution, each of the $k$ queried layers can only be queried \emph{in the same operation block}.
Therefore, at different queries to the same layer, the graph remains fixed. 
Let us consider the following $B+1$ answer matrices $A_0,A_1,\ldots,A_B$.
Each matrix is associated to a layer, and of size $B\times k$.
Each row of the matrix corresponds to a node in this layer, and each column corresponds to a queried layer.
The value in entry $A_{i,j,l}$ indicates whether node $v_{i,j}$ is connected to the top half of the $l$-th queried layer, in the graph \emph{at the time of the query to this queried layer}.
Basically, these $B+1$ matrices store the answers to all possible queries that could appear in $I_B$.
We will state the sufficient condition in terms of these matrices, and we will first give two technical definitions below.

Let us fix a (small) subset $\cL$ of the queried layers.
$\cL$ induces a natural partition of the vertices into $2^{|\cL|}$ equivalent classes: 
restrict each matrix to the columns corresponding to $\cL$, there are $2^{|\cL|}$ possible assignments to each row (node); two nodes are in the same equivalent class if the two corresponding rows in their matrices are identical.
On average, each equivalent class contains $B/2^{|\cL|}$ nodes from each layer.
We say the sequence of $k$ missing matchings is \emph{regular} for this game $G(v, O, Q_A, U_B)$ if this is ``approximately true'' for every small subset $\cL$ and every layer.
\begin{definition}
The sequence of $k$ missing matchings is \emph{regular} for game $G(v, O, Q_A, U_B)$ if for every subset $\cL$ of queried layers with $|\cL|\leq \frac{1}{2}\log B$, every equivalent class induced by $\cL$ and every layer, the size of the intersection between the layer and the equivalent class is at most $2B/2^{|\cL|}$.
\end{definition}

It is easy to show by concentration, a uniformly random sequence of $k$ missing matchings is regular with extremely high probability.
\begin{lemma}\label{lem_regular}
A uniformly random sequence of $k$ missing matchings is regular with probability at least $1-\exp(-B^{1/2})$.
\end{lemma}
\begin{proof}
Fix a set of queried layers $\cL$ with $|\cL|\leq \frac{1}{2}\log B$, one equivalent class induced by $\cL$ with assignment $c_l$ in the $l$-th entry of the row for $l\in \cL$ and one layer $i$.
Let $X_j$ be the indicator variable for the event that $v_{i,j}$ is the equivalent class, i.e., $A_{i,j,l}=c_l$ for all $l\in\cL$.
Since there is at least one missing layer between any two queried layers, for any set $S\subset \cL$, we have
\[
\begin{aligned}
	\Pr[\wedge_{j\in S}X_j=1]&\leq \left(\frac{B/2}{B}\cdot\frac{B/2-1}{B-1}\cdot\cdots\cdot\frac{B/2-|S|+1}{B-|S|+1}\right)^{|\cL|} \\
	&\leq 2^{-|S|\cdot|\cL|}.
\end{aligned}
\]
Thus, by Theorem~\ref{thm:gen_chernoff} and Remark~\ref{rem_chernoff} (setting $\mu_i=2^{-|\cL|}$ and $\alpha=1$)
\[
	\Pr[\sum_{j=1}^B X_j>2B/2^{|\cL|}]\leq 2^{-\frac{1}{2}B/2^{|\cL|}}\leq 2^{-\frac{1}{2}B^{1/2}},
\]
i.e., the probability that the size of intersection between this equivalent class and $V_i$ i more than $2B/2^{|\cL|}$ is at most $\exp(-B^{1/2})$.

Finally, by union bound and the fact that there are at most $\exp(\log^2 B)$ small sets $\cL$, $\exp(\log B)$ equivalent classes per set, and $\exp(\log B)$ layers, we prove the lemma.

\end{proof}

Before any communication, Bob starts with most entries in these $B+1$ matrices unknown.
Each time when Bob learns the answer to one query, he will be able to fill in about $B/k$ entries in these matrices directly.
That is, Bob knows the graph except the missing matchings at the time of the query.
There are $B/k$ layers between any two missing matchings.
By the transitivity of connectivity, Bob will be able to fill in one entry in each of the $B/k$ answer matrices (unless the node is before the first missing matching or after the last, in which case Bob may only fill in fewer entries).

We say a set of queries $Q$ is \emph{evenly-spreading} if $Q$ does not contain any query on the left of the leftmost missing layer, and when Bob fills in the entry values according to the above procedure for all queries in $Q$, no entry is filled in for more than once (no two queries reveal exactly the same bit) and in all matrices associated to a layer \emph{on the immediate right side of a missing matching}, each row has no more than $\frac{1}{2}\log B$ entries filled in (not too many queries reveal bits about the same node).
Formally, we define \emph{evenly-spreading} as follows.
\begin{definition}
Let $Q$ be a set of queries.
For every vertex $u$ that is in the immediate right layer of a missing matching (the layer with the larger index), let $Q_u$ be the set of all queries $q=(v, L)$ in $Q$ such that in the graph when layer $L$ is queried, $u$ and $v$ are connected without using edges from the missing matchings, i.e., Bob knows $u$ and $v$ are connected and the answer to $q$ directly translates to one answer to $(u, L)$.
We say $Q$ is \emph{evenly-spreading} if no query $(v, L)\in Q$ has $v$ on the left of the leftmost missing matching, and for each $Q_u$,
\begin{itemize}
	\item $Q_u$ does not contain two queries $(v, L)$ with the same $L$, and
	\item $|Q_u|\leq \frac{1}{2}\log B$.
\end{itemize}
\end{definition}
Note that the definition of evenly-spreading does not depend on the missing matchings.
The following lemma shows that it is extremely likely to find evenly-spreading subset from a large random set of queries.

\begin{lemma}\label{lem_even_spd}
Let $Q^{(1)},\ldots,Q^{(t)}$ be $t$ independent samples of Bob's inputs.
Then the multi-set $Q=\cup_{i=1}^t Q^{(i)}$ has an evenly-spreading subset of size $(1-\gamma/2)tkB$ with probability at least
\[
	1-2^{-\frac{1}{18}\gamma tkB\log\frac{\log B}{2t}}
\]
as long as $t\leq \frac{1}{4} \log B$, $k\geq B^{1/8}$, $\gamma>1/\log B$ and $B$ is sufficiently large.
\end{lemma}
\begin{proof}
Similar to the above definition, for every vertex $u$ that is on the immediate right layer of a missing matching, let $Q_{u}$ be the multi-set of all queries $q=(v, L)$ in $Q$ such that in the graph when layer $L$ is queried, $u$ and $v$ are connected without using edges from the missing matchings.

The largest evenly-spreading subset of $Q$ can be obtained by taking the union of the largest subset $Q'_u$ of each $Q_u$, such that
\begin{enumerate}
	\item $Q'_u$ does not contain two queries $(v, L)$ with the same $L$, and
	\item $|Q'_u|\leq \frac{1}{2}\log B$.
\end{enumerate}
Thus, it suffices to prove $\sum_u |Q'_u|\leq (1-\gamma/2)tkB$ with extremely small probability.
Since $\sum_u |Q_u|\geq (1-1/k)tkB$, 
$\sum_u |Q'_u| \leq (1-\gamma/2)tkB$ would imply that either
\begin{enumerate}
	\item at least $\gamma tkB/4-tB$ queries $(v, L)$ belong to the same $Q'_u$ with some other query with the same $L$, or
	\item $\sum_u \max\{|Q_u|-\frac{1}{2}\log B, 0\}\geq \gamma tkB/4$.
\end{enumerate}

We will show either case happens with very small probability, and by union bound, we have the desired probability.
The probability that the first case happens is at most
\[
\begin{aligned}
	\binom{k^2B}{\gamma tkB/4-tB}\cdot \left(\binom{tkB/k^2}{2}\cdot \frac{1}{B^2}\right)^{\gamma tkB/4-tB} &\leq (4ek/\gamma t)^{\gamma tkB/4-tB}\cdot \left(t/k\right)^{\gamma tkB/2-2tB} \\
	&=\left(4e t/\gamma k\right)^{\gamma tkB/4-tB} \\
	&=2^{-(\frac{1}{4}\gamma k-1) tB\cdot\log \frac{\gamma k}{4et}} \\
	&\leq 2^{-\frac{1}{40}\gamma ktB\cdot \log\frac{B}{t}}.
\end{aligned}
\]

By Chernoff bound and union bound, the probability that the second case happens is at most
\[
\begin{aligned}
	&\Pr\left[\exists C, \forall u\in C,|Q_u|\geq \frac{1}{2}\log B, \sum_{u\in C}|Q_u|\geq \frac{1}{2}|C|\log B+\gamma tkB/4\right] \\
	&\leq\sum_{c=1}^{kB}\binom{kB}{c}\cdot 2^{-\frac{1}{2}(\frac{1}{2}c\log B+\frac{1}{4}\gamma tkB-ct)\log \frac{\frac{1}{2}c\log B+\frac{1}{4}\gamma tkB}{ct}} \\
	&\leq \sum_{c=1}^{\frac{1}{2}\gamma tkB/\log B}\binom{kB}{c}\cdot 2^{-\frac{1}{8}\gamma tkB\cdot \log\frac{\log B}{2t}} \\
	&+\sum_{c=\frac{1}{2}\gamma tkB/\log B+1}^{kB}\binom{kB}{c}\cdot 2^{-\frac{1}{8}c\log B\log \frac{\log B}{2t}}\\
	&\leq \left(\frac{ekB}{\gamma tkB/2\log B}\right)^{\gamma tkB/\log B}\cdot 2^{-\frac{1}{8}\gamma tkB\cdot \log\frac{\log B}{2t}} \\
	&+\sum_{c\geq \frac{1}{2}\gamma tkB/\log B+1}\left(\frac{ekB}{\gamma tkB/2\log B}\right)^c\cdot 2^{-\frac{1}{8}c\log B\cdot \log \frac{\log B}{2t}} \\
	&=\sum_{c\geq \frac{1}{2}\gamma tkB/\log B}2^{c\log \left(\frac{2e\log B}{\gamma t}\right)}\cdot 2^{-\frac{1}{8}c\log B\cdot \log \frac{\log B}{2t}} \\
	&\leq \sum_{c\geq \frac{1}{2}\gamma tkB/\log B} 2^{-(\frac{1}{8}-o(1))c\log B\cdot \log \frac{\log B}{2t}} \\
	&\leq 2^{-\frac{1}{17}\gamma tkB\log\frac{\log B}{2t}}.
\end{aligned}
\]
Finally, by union bound, we prove the desired result.
\end{proof}

The following lemma asserts that it is a sufficient condition that the sequence of $k$ missing matchings is regular and the set of queries $Q$ is evenly-spreading for each query to reveal about one bit of information.
\begin{lemma}\label{lem_enc_perm}
For any set of \emph{evenly-spreading} queries $Q$, given the answers to $Q$, there is an encoding scheme for the $k$ missing matchings such that any regular sequence of missing matchings is encoded in $2kB+kB\log B-|Q|$ bits.
\end{lemma}
\begin{proof} We first present the encoding scheme.
\paragraph{Encoding.} Fix queries $Q$, given the $k$ missing matchings, one encodes them as follows.
\begin{enumerate}
	\item Write down $A_0$;
	\item For each vertex $u$ that is in the immediate right layer of a missing matching, find the set $M_u$ of vertices $u'$ on the left of the same missing matching, such that for all queries $q=(v, L)\in Q_u$, $(u, L)$ and $(u', L)$ have the same answer, sort $M_u$ alphabetically, write down the index of the vertex in this sorted list that $u$ is connected to.
\end{enumerate}
\paragraph{Decoding.} Next, we show that given the answers to $Q$ and the encoding, one can reconstruct the missing matching.
\begin{enumerate}
	\item Read $A_0$ from the encoding;
	\item For $i=0$ to $B$, derive $A_{i+1}$ from $A_i$ (recall that $A_i$ is the $B\times k$ 0-1 matrix storing all answers to queries that could be asked in $I_B$ and of form $(u, L)$ where $u$ is in layer $i$):
	\begin{enumerate}[i)]
		\item if the matching between layer $i$ and $i+1$ is not updated in either $I_A$ or $I_B$, then this matching is known (hardwired in game $G$), and $A_{i+1}$ can be obtained from $A_i$ by permuting the rows according to this matching (see Figure~\ref{fig:mat1});
		\item if the matching between the two layers is a replaced matching, then the matchings before and after the replacement are both known,\footnote{The matching before the replacement is always the identity matching.} and the time of the replacement is also known, $A_{i+1}$ can be obtained by permuting the columns corresponding to a query that happens after the replacement, according the new matching (see Figure~\ref{fig:mat2});
		\item if the matching between the two layers is a missing matching, go over all $u$ in layer $i+1$, find and sort $M_u$,\footnote{This can be done since we already know $A_i$, and we also know the answers to all queries $q=(v, L)\in Q_u$, which has the same answer as $(u, L)$.} then find the vertex that $u$ is connected to from the encoding, which recovers the missing matching and $A_{i+1}$ can be obtained by permuting the rows of $A_i$ (see Figure~\ref{fig:mat3}).
	\end{enumerate}
\end{enumerate}
\begin{figure}
\centering
\begin{minipage}[t]{0.27\linewidth}
\centering
\begin{tikzpicture}
  \foreach \i in {1,2,3,4} {
    \node [draw,circle,inner sep=0pt,minimum size=3pt] at (0 pt,\i*25 pt) (u_\i) {};
    \node [draw,circle,inner sep=0pt,minimum size=3pt] at (30 pt,\i*25 pt) (v_\i) {};
  }
  \foreach \i/\j/\c/\cl in {1/2/010110/black!50,2/4/101011/black,3/3/111101/black!50,4/1/010000/black!50} {
    \draw [thick] (u_\i) -- (v_\j);
    \node [text=\cl] at (-30pt, \i*25 pt) (t_\i) {\bf\small \c};
    \node [text=\cl] at (60pt, \j*25 pt) {\bf\small \c};
  }
  \node at (-30pt, 0pt) {$\mathbf{A_i}$};
  \node at (60pt, 0pt) {$\mathbf{A_{i+1}}$};
\end{tikzpicture}
\caption{Case i) fixed matching.}\label{fig:mat1}
\end{minipage}
\hspace{20pt}
\begin{minipage}[t]{0.27\linewidth}
\centering
\begin{tikzpicture}
  \foreach \i in {1,2,3,4} {
    \node [draw,circle,inner sep=0pt,minimum size=3pt] at (0 pt,\i*25 pt) (u_\i) {};
    \node [draw,circle,inner sep=0pt,minimum size=3pt] at (30 pt,\i*25 pt) (v_\i) {};
  }
  \foreach \i/\j in {1/2,2/4,3/3,4/1}{
    \draw [thick] (u_\i) -- (v_\j);
    \draw [dashed] (u_\i) -- (v_\i);
  }
  \node at (-30pt, 25pt) {\small \bf\color{black!50}010110};
  \node at (-30pt, 50pt) {\small \bf 101{\color{red}011}};
  \node at (-30pt, 75pt) {\small \bf\color{black!50}111101};
  \node at (-30pt, 100pt) {\small \bf\color{black!50}010000};

  \node at (60pt, 25pt) {\small \bf\color{black!50}010000};
  \node at (60pt, 50pt) {\small \bf{\bf 101}\color{black!50}110};
  \node at (60pt, 75pt) {\small \bf\color{black!50}111101};
  \node at (60pt, 100pt) {\small \bf{\color{black!50}010}\bf\color{red}011};
  \node at (-30pt, 0pt) {$\mathbf{A_i}$};
  \node at (60pt, 0pt) {$\mathbf{A_{i+1}}$};
\end{tikzpicture}
\caption{Case ii) replaced matching. First three entries are before the update, last three entries are after the update.}\label{fig:mat2}
\end{minipage}
\hspace{20pt}
\begin{minipage}[t]{0.27\linewidth}
\centering
\begin{tikzpicture}
  \foreach \i in {1,2,3,4} {
    \node [draw,circle,inner sep=0pt,minimum size=3pt] at (0 pt,\i*25 pt) (u_\i) {};
    \node [draw,circle,inner sep=0pt,minimum size=3pt] at (30 pt,\i*25 pt) (v_\i) {};
  }
  \node at (-30pt, 25pt) {\small \bf 010110};
  \node at (-30pt, 50pt) {\small \bf 101011};
  \node at (-30pt, 75pt) {\small \bf 111101};
  \node at (-30pt, 100pt) {\small \bf 010000};

  \node at (60pt, 25pt) {\small \bf ?????0};
  \node at (60pt, 50pt) {\small \bf ??0?1?};
  \node at (60pt, 75pt) {\small \bf ?1????};
  \node at (60pt, 100pt) {\small \bf 1????1};

  \node at (-30pt, 0pt) {$\mathbf{A_i}$};
  \node at (60pt, 0pt) {$\mathbf{A_{i+1}}$};

  \foreach \i/\j in {1/2,2/4,3/3,4/1,1/1,4/3,3/4}{
    \draw [dotted] (u_\i) -- (v_\j);
  }
\end{tikzpicture}
\caption{Case iii) missing matching. All possible matching edges are listed.}\label{fig:mat3}
\end{minipage}
\end{figure}
\paragraph{Analysis.}
Since the encoding explicitly writes down for each $u$ adjacent to a missing matching, among all possible neighbors of $u$ based on the answers to $Q$, the vertex it is connected to, the decoding procedure will successfully find its neighbor, which recovers the missing matchings.

The first part $A_0$ costs $kB$ bits.
When the sequence of $k$ missing matchings is regular, since $Q$ is evenly-spreading, it costs $\log (2B/2^{|Q_u|})=\log B+1-|Q_u|$ bits to specify the neighbor of $u$ on the other side of the missing matching.
We also have $\sum |Q_u|=|Q|$.
Thus, the total encoding length is
\[
	kB+\sum_{u} (\log B+1-|Q_u|)=2kB+kB\log B-|Q|.
\]
This proves the lemma.
\end{proof}

\paragraph{Updates in the hard distribution $\cD$.}
Before proving the communication lower bound, let us first finish describing the updates in the hard distribution.
The goal here is to design a distribution which satisfies the property in Proposition~\ref{prop_dist} and is supported on regular inputs for every game (in order to apply Lemma~\ref{lem_enc_perm}).
We begin with the uniform distribution over all $(B!)^B$ sequences of new matchings conditioned on \emph{Alice's inputs being regular in all $\Theta(B)$ communication games}.
Denote this distribution by $\cD_{\mathrm{reg}}$.
By union bound, a uniformly random sequence of updates induces regular inputs for all $\Theta(B)$ games with probability at least $1-\exp(-B^{1/2})>1/2$.
Thus, $\cD_{\mathrm{reg}}$ has support size at least $\frac{1}{2}(B!)^B$.

Next, we refine $\cD_{\mathrm{reg}}$ so that the distribution will satisfy the property in Proposition~\ref{prop_dist}.
The refinement has $B$ rounds.
Denote the distribution after round $i$ by $\cD_i$, and $\cD_0=\cD_{\mathrm{reg}}$.
Denote by $M_l$ the random variable indicating $l$-th updated matching in chronological order.
We will show by induction that $\cD_i$ has the following property:
for all $0\leq j\leq i$ and any $m_1,\ldots,m_j$ in the support of $\cD_i$, the distribution for the remaining $B-j$ matchings conditioned on $M_l=m_l$ for all $1\leq l\leq j$ is a uniform distribution with support size at least $2^{-i-1} (B!)^{B-j}$.
It is easy to verify that
\begin{itemize}
	\item $\cD_0$ satisfies this property, and
	\item if $\cD_B$ satisfies this property, then by setting $\cD=\cD_B$, Proposition~\ref{prop_dist} follows.
\end{itemize}

It remains to show how to do each round of the refinement and complete the induction step.
In round $i$, we go over all possible values $m_1,\ldots,m_{i+1}$.
If the remaining $B-(i+1)$ matchings have support size smaller than $2^{-(i+1)-1} (B!)^{B-(i+1)}$ conditioned on $M_j=m_j$ for all $1\leq j\leq i+1$, then we remove all sequences that start with $(m_1,\ldots,m_{i+1})$ from the support.
Let the new uniform distribution be $\cD_{i+1}$.
By definition, for $j=i+1$ and any $m_1,\ldots,m_j$ in the support of $\cD_{i+1}$, the support size of the remaining $B-j$ matchings conditioned on these $j$ matchings is at least $2^{-(i+1)-1}(B!)^{B-(i+1)}$, which satisfies the property.
For any $0\leq j\leq i$, and $m_1,\ldots,m_j$ in the support, the support size of the remaining $B-j$ matchings conditioned on these $j$ matchings is reduced by at most
\[
	(B!)^{i-j+1}\cdot 2^{-(i+1)-1} (B!)^{B-(i+1)}=2^{-(i+1)-1}(B!)^{B-j}.
\]
However, by induction hypothesis, the support size in $\cD_i$ which we begin with is $2^{-i-1}(B!)^{B-j}$.
Thus, when $0\leq j\leq i$, the support size is also at least $2^{-(i+1)-1}(B!)^{B-j}$.
This proves Proposition~\ref{prop_dist}.

Now we are ready to prove Lemma~\ref{lem_dcon_all}.

\begin{proof}[Proof of Lemma~\ref{lem_dcon_all}]

Assume for contraction, there is a a too-efficient communication protocol $P$ for game $G(v, O, Q_A, U_B)$, where Alice's first message length is $(1+\epsilon_v\log B) kB+O(\log B)$, and Bob answers all queries correctly with probability at least $2^{-(1-\gamma) kB}$.
By Markov's inequality, for at least $2^{-(1-\gamma) kB-1}$-fraction of Alice's inputs $X$, the probability that Bob answers all queries in $S$ correctly with probability at least $2^{-(1-\gamma) kB-1}$ conditioned on Alice's input being $X$.
Denote this subset of Alice's input by $\cX$.
We have $\Pr_{X}[X\in \cX]\geq 2^{-(1-\gamma) kB-1}$.
Since the min-entropy of $X$ is at least $k\log B!-2kB$, we have
\[
	\log |\cX|\geq k\log B!-2kB-(1-\gamma) kB-1> kB\log B-5kB.
\]
We are going to design a too-efficient encoding scheme for $\cX$ using this hypothetical protocol, assuming there is a shared random string between the encoder and the decoder.

\paragraph{Encoding.} Given a sequence of $k$ missing matchings from $\cX$ and a shared random string, we are going to encode it as follows.
\begin{enumerate}
	\item Simulate $P$ as Alice, assuming the $k$ missing matchings are the input. Write down Alice's first message to Bob.
	\item View the shared random string as infinite samples of sequences of queries in $I_B$ (according to the input distribution). 
	Divide the them into chunks of $t=\frac{4\epsilon_v}{\gamma}\log n$ samples each, and index chunks by natural numbers.
	Write down the index of the first chunk such that for all samples in the chunk, Bob will be able to answer all queries correctly based on Alice's first message. (Note that this number can be unbounded, and such a number $x$ can be encoded in $\log x + O(\log\log x)$ bits.)
	\item Let $Q$ be the largest evenly-spreading subset of the set of all queries in this chunk, in case of a tie, let $Q$ be the lexicographically first one.
	Encode the missing matchings using the encoding scheme described in Lemma~\ref{lem_enc_perm}.
\end{enumerate}
\paragraph{Decoding.} The following decoding procedure recovers the missing matchings.
\begin{enumerate}
	\item Simulate $P$ as Bob, read Alice's first message.
	\item Read the index of the first all-correct chunk. For each sample in the chunk, simulate $P$ as Bob to answer all the queries in it.
	\item Find $Q$ and use the decoding procedure in Lemma~\ref{lem_enc_perm} to reconstruct the missing matchings given the answers to $Q$.
\end{enumerate}

\paragraph{Analysis.} By Lemma~\ref{lem_enc_perm}, the decoding procedure above successfully reconstructs the missing matchings, since the sequence of missing matchings is regular, $Q$ is evenly-spreading and all queries in $Q$ are answered correctly.

Now let us analyze the number of bits used in the encoding procedure.
In Step 1, it takes $(1+\epsilon_v\log B) kB+1$ bits to write down Alice's message.
In Step 2, each sample is correct with probability at least $2^{-(1-\gamma) kB-1}$.
Thus, each chunk is all correct with probability at least $2^{-((1-\gamma)kB+1)t}$.
Thus, the expected index of the first all-correct chunk is at most $2^{((1-\gamma) kB+1)t}$.
By concavity of the logarithm, it takes at most $((1-\gamma) kB+1)t+O(\log B)$ bits to write down this index in expectation.
By Lemma~\ref{lem_enc_perm}, Step 3 takes $2kB+kB\log B-|Q|$ bits.

On the other hand, $t=\frac{\epsilon_v}{6\log (1/\epsilon_v)}<\frac{1}{4}\log B$ and $\gamma>1/\log B$.
By Lemma~\ref{lem_even_spd} and the fact that $\Pr[W|E]\leq\Pr[W]/\Pr[E]$, we have 
\[
\begin{aligned}
	\Pr[|Q|<(1-\gamma/2)tkB\mid \textrm{the chunk is all correct}]&\leq 2^{-\frac{1}{18}\gamma tkB\log\frac{\log B}{2t}}/2^{(-(1-\gamma) kB-1)t}\\
	&\leq 2^{tkB(1-\frac{1}{18}\gamma \log\frac{\log B}{2t})}.
\end{aligned}
\]
Since $\gamma=24/\log (1/\epsilon_v)$, we have
\[
\begin{aligned}
	\gamma\log\frac{\log B}{2t}&=\gamma\log \frac{\gamma}{8\epsilon_v} \\
	&=\frac{24}{\log (1/\epsilon_v)}\log \frac{3}{\epsilon_v\log (1/\epsilon_v)} \\
	&>20
\end{aligned}
\]
for $\epsilon_v<1$.
Thus, the probability that $|Q|<(1-\gamma/2)tkB$ is at most $2^{-\frac{1}{9}tkB}$.
In expectation, Step 3 takes at most 
\[
	2kB+kB\log B-\E[|Q|]\leq 2kB+kB\log B-(1-2^{-\frac{1}{9}tkB})(1-\gamma/2)tkB.
\]

Finally, summing up all three steps, the expected total encoding length is at most
\[
	\begin{aligned}
		&(1+\epsilon_v\log B) kB+((1-\gamma) kB+1)t+O(\log B)+2kB+kB\log B-(1-2^{-\frac{1}{9}tkB})(1-\gamma/2)tkB \\
		&\leq\epsilon_v kB\log B+(1-\gamma)tkB+O(\log B)+3kB+kB\log B-(1-\gamma/2-2^{-\frac{1}{9}tkB})tkB \\
		&\leq\epsilon_v kB\log B-\gamma tkB/2+O(\log B)+3kB+kB\log B \\
		&\leq-\epsilon_v kB\log B+O(\log B)+3kB+kB\log B \\
		&\leq kB\log B-6 kB+O(\log B) \\
		&<\log |\cX|
	\end{aligned}
\]
for sufficiently large $B$.

Thus, there is a way to fix the public random string such that a uniformly random element from $\cX$ can be encoded using strictly fewer than $\log |\cX|$ bits in expectation, yielding a contradiction.
\end{proof}

\ifx\undefined\mainfile
\bibliography{matrix-seq}
\bibliographystyle{alpha}
\end{document}
\fi

\section*{Acknowledgment}
We would like to thank Pritish Kamath for pointing out an improvement to our original online set intersection protocol.

\bibliographystyle{alpha}
\bibliography{matrix-seq}

\begin{thebibliography}{MNSW95}

\bibitem[Ajt88]{Ajtai88}
Mikl{\'{o}}s Ajtai.
\newblock A lower bound for finding predecessors in yao's cell probe model.
\newblock {\em Combinatorica}, 8(3):235--247, 1988.

\bibitem[BF02]{BF02}
Paul Beame and Faith~E. Fich.
\newblock Optimal bounds for the predecessor problem and related problems.
\newblock {\em J. Comput. Syst. Sci.}, 65(1):38--72, 2002.

\bibitem[BH11]{TM3}
George~F Burkhard and Eric~T Hoke.
\newblock Transfer matrix optical modeling.
\newblock 2011.

\bibitem[Blu85]{Blum85}
Norbert Blum.
\newblock On the single-operation worst-case time complexity on the disjoint
  set union problem.
\newblock In {\em {STACS} 85, 2nd Symposium of Theoretical Aspects of Computer
  Science, Saarbr{\"{u}}cken, Germany, January 3-5, 1985, Proceedings}, pages
  32--38, 1985.

\bibitem[CGL15]{CGL15}
Rapha{\"{e}}l Clifford, Allan Gr{\o}nlund, and Kasper~Green Larsen.
\newblock New unconditional hardness results for dynamic and online problems.
\newblock In {\em {IEEE} 56th Annual Symposium on Foundations of Computer
  Science, {FOCS} 2015, Berkeley, CA, USA, 17-20 October, 2015}, pages
  1089--1107, 2015.

\bibitem[CP10]{chattopadhyay2010story}
Arkadev Chattopadhyay and Toniann Pitassi.
\newblock The story of set disjointness.
\newblock {\em ACM SIGACT News}, 41(3):59--85, 2010.

\bibitem[DF04]{dummit2004abstract}
David~Steven Dummit and Richard~M Foote.
\newblock {\em Abstract algebra}, volume~3.
\newblock Wiley Hoboken, 2004.

\bibitem[FS89]{FS89}
Michael~L. Fredman and Michael~E. Saks.
\newblock The cell probe complexity of dynamic data structures.
\newblock In {\em Proceedings of the 21st Annual {ACM} Symposium on Theory of
  Computing, May 14-17, 1989, Seattle, Washigton, {USA}}, pages 345--354, 1989.

\bibitem[HK99]{henzinger1999randomized}
Monika~R Henzinger and Valerie King.
\newblock Randomized fully dynamic graph algorithms with polylogarithmic time
  per operation.
\newblock {\em Journal of the ACM (JACM)}, 46(4):502--516, 1999.

\bibitem[Hoe63]{Hoeffding63}
Wassily Hoeffding.
\newblock Probability inequalities for sums of bounded random variables.
\newblock {\em Journal of the American statistical association},
  58(301):13--30, 1963.

\bibitem[HW07]{HW07}
Johan H{\aa}stad and Avi Wigderson.
\newblock The randomized communication complexity of set disjointness.
\newblock {\em Theory of Computing}, 3(1):211--219, 2007.

\bibitem[IK10]{IK10}
Russell Impagliazzo and Valentine Kabanets.
\newblock {\em Constructive Proofs of Concentration Bounds}, pages 617--631.
\newblock 2010.

\bibitem[LD69]{LinDonaldson}
YK~Lin and BK~Donaldson.
\newblock A brief survey of transfer matrix techniques with special reference
  to the analysis of aircraft panels.
\newblock {\em Journal of Sound and Vibration}, 10(1):103--143, 1969.

\bibitem[LW17]{LW17}
Kasper~Green Larsen and R.~Ryan Williams.
\newblock Faster online matrix-vector multiplication.
\newblock In {\em SODA}, pages 2182--2189, 2017.

\bibitem[MNSW95]{MNSW95}
Peter~Bro Miltersen, Noam Nisan, Shmuel Safra, and Avi Wigderson.
\newblock On data structures and asymmetric communication complexity.
\newblock In {\em STOC}, pages 103--111, 1995.

\bibitem[P{\v a}t07]{Patrascu07}
Mihai P{\v a}tra{\c s}cu.
\newblock Lower bounds for 2-dimensional range counting.
\newblock In {\em STOC}, pages 40--46, 2007.

\bibitem[PD04a]{PD04}
Mihai P{\v{a}}tra{\c{s}}cu and Erik~D Demaine.
\newblock Lower bounds for dynamic connectivity.
\newblock In {\em STOC}, pages 546--553, 2004.

\bibitem[PD04b]{puaatracscu2004tight}
Mihai P{\u{a}}tra{\c{s}}cu and Erik~D Demaine.
\newblock Tight bounds for the partial-sums problem.
\newblock In {\em SODA}, pages 20--29, 2004.

\bibitem[PD06]{PD06}
Mihai P{\u{a}}tra{\c{s}}cu and Erik~D Demaine.
\newblock Logarithmic lower bounds in the cell-probe model.
\newblock {\em SIAM Journal on Computing}, 35(4):932--963, 2006.

\bibitem[PRI99]{TM1}
Leif~AA Pettersson, Lucimara~S Roman, and Olle Ingan{\"a}s.
\newblock Modeling photocurrent action spectra of photovoltaic devices based on
  organic thin films.
\newblock {\em Journal of Applied Physics}, 86(1):487--496, 1999.

\bibitem[PS97]{PS97}
Alessandro Panconesi and Aravind Srinivasan.
\newblock Randomized distributed edge coloring via an extension of the
  chernoff--hoeffding bounds.
\newblock {\em SIAM J. Comput.}, 26(2):350--368, April 1997.

\bibitem[PT11]{PT11}
Mihai P{\u{a}}tra{\c{s}}cu and Mikkel Thorup.
\newblock Don't rush into a union: take time to find your roots.
\newblock In {\em STOC}, pages 559--568, 2011.

\bibitem[PYF03]{TM2}
Peter Peumans, Aharon Yakimov, and Stephen~R Forrest.
\newblock Small molecular weight organic thin-film photodetectors and solar
  cells.
\newblock {\em Journal of Applied Physics}, 93(7):3693--3723, 2003.

\bibitem[RR16]{RR16}
Sivaramakrishnan~Natarajan Ramamoorthy and Anup Rao.
\newblock Simplified data structure lower bounds for dynamic graph
  connectivity.
\newblock {\em Electronic Colloquium on Computational Complexity {(ECCC)}},
  23:167, 2016.

\bibitem[She14]{sherstov2014communication}
Alexander~A Sherstov.
\newblock Communication complexity theory: Thirty-five years of set
  disjointness.
\newblock In {\em International Symposium on Mathematical Foundations of
  Computer Science}, pages 24--43. Springer, 2014.

\bibitem[Smi90]{Smid90}
Michiel H.~M. Smid.
\newblock A data structure for the union-find problem having good
  single-operation complexity.
\newblock {\em ALCOM: Algorithms Review, Newsletter of the ESPRIT II Basic
  Research Actions Program}, 1990.

\bibitem[ST81]{sleator1981}
Daniel~D Sleator and Robert~Endre Tarjan.
\newblock A data structure for dynamic trees.
\newblock In {\em Proceedings of the thirteenth annual ACM symposium on Theory
  of computing}, pages 114--122. ACM, 1981.

\bibitem[Tho00]{thorup2000}
Mikkel Thorup.
\newblock Near-optimal fully-dynamic graph connectivity.
\newblock In {\em Proceedings of the thirty-second annual ACM symposium on
  Theory of computing}, pages 343--350. ACM, 2000.

\bibitem[WY16]{WY16}
Omri Weinstein and Huacheng Yu.
\newblock Amortized dynamic cell-probe lower bounds from four-party
  communication.
\newblock In {\em FOCS}, pages 305--314, 2016.

\bibitem[Yao77]{yao1977probabilistic}
Andrew Chi-Chin Yao.
\newblock Probabilistic computations: Toward a unified measure of complexity.
\newblock In {\em FOCS}, pages 222--227, 1977.

\bibitem[Yao81]{Yao81}
Andrew Chi-Chih Yao.
\newblock Should tables be sorted?
\newblock {\em Journal of the ACM (JACM)}, 28(3):615--628, 1981.

\bibitem[Yu16]{Yu16}
Huacheng Yu.
\newblock Cell-probe lower bounds for dynamic problems via a new communication
  model.
\newblock In {\em STOC}, pages 362--374, 2016.

\end{thebibliography}
\appendix
\ifx\undefined\mainfile





\fi

\section{Further Results about the Group Range Problem} \label{s:further-group-results}

\subsection{Groups versus Monoids}
\label{ss:monoid}

One key property of groups needed for our proof is the invertibility. Consider
generalizing to the \emph{Monoid Range Problem}, which considers general monoids
instead of groups. Monoids are sets closed under an associative operation and
have an identity element (notice they do not have the invertibility property).
We show that our lower bound does not hold for the Monoid Range Problem:

\begin{restate}[Theorem~\ref{thm:monoids}]
  There exists a family of monoids $(G_n)_n$ such that the Monoid Range Problem
  can be solved in $O \left( \frac{\log n}{\log \log n} \right)$ time per operation.
\end{restate}

\begin{proof}
  Consider the following family of monoids. We use $\times$ to denote the operator and
  $0$ to denote the identity element, and $\star$ to denote a special element. The
  family has the following property: for any elements $x, y \in G_n$ we have
  that $x \times y = \star$ unless $x$ or $y$ is $0$ (in which case their product equals the
  other, due to the identity property).

  See Table~\ref{tbl:g2g3} for small examples of these monoids. One way to think
  about these monoids is that the elements are zero, singletons, or products of more
  than one singleton ($\star$).

  \begin{table}[h]
  \centering
  \begin{tabular}{l l l}
  \begin{tabular}{| c | c c |}
    \hline
    $\times$ & 0 & $\star$ \\ \hline
    0 & 0 & $\star$ \\
    $\star$ & $\star$ & $\star$ \\ \hline
  \end{tabular}
  &
  \begin{tabular}{| c | c c c |}
    \hline
    $\times$ & 0 & 1 & $\star$ \\ \hline
    0 & 0 & 1 & $\star$ \\
    1 & 1 & $\star$ & $\star$ \\
    $\star$ & $\star$ & $\star$ & $\star$ \\ \hline
  \end{tabular}
  &
  \begin{tabular}{| c | c c c c |}
    \hline
    $\times$ & 0 & 1 & 2 & $\star$ \\ \hline
    0 & 0 & 1 & 2 & $\star$ \\
    1 & 1 & $\star$ & $\star$ & $\star$ \\
    2 & 2 & $\star$ & $\star$ & $\star$ \\
    $\star$ & $\star$ & $\star$ & $\star$ & $\star$ \\ \hline
  \end{tabular}
  \\
  \end{tabular}
  \caption{Multiplication tables for $G_2$, $G_3$, and $G_4$.}
  \label{tbl:g2g3}
  \end{table}

  Thus, the product of a sequence of elements in $G_n$ is $\star$ if there are more than one non-zero element; the product is $0$ if all elements are zeros; the product is the only if non-zero element if there is exactly one.
  To efficiently maintain the range product of a $G_n$ sequence, we use a \emph{segment tree} of branching factor $B=\Theta(\log n)$.

  \paragraph{The data structure} Assume without loss of generality, that $n$ is a power of $B$.
  Each node of the tree at depth $i$ is associated with a (contiguous) subsequence of length $n/B^i$.
  Dividing the associated subsequence of a node $E$ into $B$ subsequences evenly, the $j$-th child of $E$ is associated with the $j$-th subsequence.
  In particular, the root is associated with the entire sequence, the $j$-th child of the root is associated with $((j-1)\cdot n/B+1)$-th element to $(j\cdot n/B)$-th element, and each leaf is associated with a singleton. 
  In each node $E$ of the segment tree, the data structure maintains
  \begin{enumerate}[(1)]
  	\item
  		for each child of $E$, the minimum of two and the number of non-zero elements in their associated subsequences, i.e., if there is none or one or more than one non-zero element;
  	\item
  		if there is exactly one non-zero element in the associated subsequence of $E$, what this element is.
  \end{enumerate}
  Note that Part (1) costs $O(1)$ bits for each child, thus $O(B)=O(\log n)$ bits in total, and Part (2) costs $O(\log n)$ bits to indicate the element.
  Thus, both parts can be stored in $O(1)$ words for each node.

  \paragraph{Updates} Upon receipt of an update $a_i:=x$, the data structure iteratively updates the information in the tree bottom-up.
  It is not hard to verify that this update could only affect the nodes associated with some subsequence consisting of $i$. 
  It first finds the leaf associated with $\{i\}$, and updates the two parts according to the value of $x$.
  Once all descendants of a node $E$ are up-to-date, Part (1) of $E$ can be updated by checking Part (1) of the only child of $E$ affected by the update.
  From the updated Part (1) of $E$, one can figure out if there is exactly one non-zero element in the associated subsequence and which subtree it is in if there is. 
  By checking Part (2) of the relevant subtree, the data structure will be able to update Part (2) of $E$. 
  Updating each node takes $O(1)$ time, only $O(\log_B n)$ nodes are affected by the update.
  Thus, the total update time is $O(\log n/\log\log n)$.

  \paragraph{Queries} Recall that each child $E_i$ of the root node is associated with a subsequence of length $n/B$. 
  To answer the query $a_i \times \cdots \times a_j$, the data structure first breaks $[i, j]$ into subsequences $S_1, S_2,\ldots, S_m$, such that $S_2,\ldots,S_{m-1}$ are associated to $E_a,\ldots,E_{a+m-2}$ for some $a$, $S_1$ and $S_m$ are subsequences of the associated subsequences of $E_{a-1}$ and $E_{a+m-1}$ respectively. 
  By accessing Part (1) of the root node, the data structure learns whether there is none, exactly one or more than one non-zero elements in $S_2,\ldots,S_{m-1}$. 
  Then it recurses on $S_1$ in the subtree rooted at $E_{a-1}$ and $S_m$ in the subtree rooted at $E_{a+m-1}$. 
  By combining the answer from three parts, it will be able to output the answer to the query.
  It is not hard to verify that at each depth, at most two nodes of the tree may be recursed on. 
  The query algorithm spends $O(1)$ time in each node.
  Thus, the total query time is $O(\log n/\log\log n)$.

  Therefore, we conclude that the Monoid Range Problem with this particular family of monoids can be solved in $O(\log n/\log\log n)$ time per operation.
\end{proof}

\ifx\undefined\mainfile
\end{document}
\fi
\ifx\undefined\mainfile





\fi
\subsection{The Matrix Range Problem}
\label{s:matrix}

\newcommand{\fourmat}[4]{
\left[
\begin{array}{c|c}
#1 & #2 \\
\hline
#3 & #4
\end{array}
\right]
}
\newcommand{\stdbasis}[1][i]{\textbf{e}_{#1}}
\newcommand{\zerovec}{\textbf{0}}

In this section, we show that for a particular group $G$, even maintaining one particular bit (say the last bit) of the whole product $\prod_{i=1}^n a_i$ is hard.
The group $G$ we focus on is the general linear group of invertible
matrices over the field $\mathbb{F}_p$ for constant $p$, namely $G = GL(\sqrt{\log n}, \mathbb{F}_p)$.

The binary encoding of matrices we would like to focus on is the encoding of a matrix as the concatenation of its entries. Hence, queries will return a bit about an entry of the matrix product. We call the Group Range Problem with $G = GL(\sqrt{\log n}, \mathbb{F}_p)$ and this encoding the \emph{Matrix Range Problem}. However, since not all $\sqrt{\log n} \times \sqrt{\log n}$ matrices
over $\mathbb{F}_p$ are invertible, this is not the most concise encoding of $GL(\sqrt{\log n}, \mathbb{F}_p)$. We remark that our desired encoding is nonetheless concise enough for Theorem~\ref{thm:group_range_LB} to hold:

\begin{lemma}
Theorem~\ref{thm:group_range_LB} holds for the Matrix Range Problem.
\end{lemma}


\begin{proof}
Consider the group of $\sqrt{\log n} \times \sqrt{\log n}$ invertible matrices
over the field $\mathbb{F}_p$ where $p$ is constant. Recall that this group has
$|GL(\sqrt{\log n}, p)| = \prod_{i=0}^{\sqrt{\log n}-1} (p^n - p^i)$ elements (see e.g. \cite[page 413]{dummit2004abstract}).
We would like to represent this group in usual matrix format, i.e. as the
concatenation of the representations of their entries. This representation uses
$\log n \log p$ bits. On the other hand, notice that
$\log |GL(n, p)| \ge (\log n - \sqrt{\log n}) \log p$, so
Theorem~\ref{thm:group_range_LB} implies that our lower bound holds for
this setting.
\end{proof}

The \emph{Matrix Product Problem} is the same as
the Matrix Range Problem, except that instead of being able to query for (a bit of) any entry of the product of the matrices in any subinterval, we are only allowed to query for (a bit of) the bottom-right entry of the product of the entire range of matrices. 
Despite this substantial restriction on the types of queries allowed, we find that the \emph{Matrix Range Problem} can be reduced to the \emph{Matrix Product Problem} such that our lower bounds from the previous section still apply to the \emph{Matrix Product Problem}.

\begin{lemma} \label{lem:mat-prod}
If the Matrix Product Problem for $n$ matrices of dimension $d \times d$ can be solved in amortized $T(n,d)$ time per operation, then the Matrix Range Problem $n$ matrices of dimension $d \times d$ can be solved in amortized $O(T(n,d+1))$ time per operation.
\end{lemma}

\begin{proof}
The inspiration for the reduction is the following fact: Let $\stdbasis{}$ denote the $i^{th}$ standard basis vector, i.e. the vector of length $n$ whose entries are all 0 except for its $i^{th}$ entry which is 1. For any $d \times d$ matrices $A$, $B$, and $C$, consider the following product of three $(d+1) \times (d+1)$ matrices:
\[
D =
  \fourmat{A}{\zerovec{}}{\stdbasis[j']^T}{1}
  \fourmat{B}{\zerovec{}}{\zerovec{}^T}{1}
  \fourmat{C}{\stdbasis[i']}{\zerovec{}^T}{1}
\]

In the resulting matrix $D$, the bottom right entry $D_{(d+1)(d+1)}$ is equal to $(B_{i'j'}+1)$.

The reduction is hence as follows. For any sequence $M_1, M_2, \ldots, M_n$ of $d \times d$ matrices, and any two indices $1 < i < j < n$, consider the following product of $(d+1) \times (d+1)$ matrices:

\[
D =
\fourmat{M_1}{\zerovec{}}{\zerovec{}^T}{1}
\fourmat{M_2}{\zerovec{}}{\zerovec{}^T}{1}
\cdots
\fourmat{M_{i-1}}{\zerovec{}}{\stdbasis[j']^T}{1}
\fourmat{M_i}{\zerovec{}}{\zerovec{}^T}{1}
\cdots
\fourmat{M_j}{\zerovec{}}{\zerovec{}^T}{1}
\fourmat{M_{j+1}}{\stdbasis[i']}{\zerovec{}^T}{1}
\cdots
\fourmat{M_n}{\zerovec{}}{\zerovec{}^T}{1}
\]

Similar to before, the bottom right entry $D_{(d+1)(d+1)}$ will be equal to the $(i',j')^{th}$ entry of the product $M_i \cdots M_j$ plus one. To deal with $i = 1$, then no matrix has $\stdbasis[j']^T$ as its bottom row. The first $d$ entries of the right column of $D$ will be the $i'^{th}$ column of $M_1 \cdots M_j$. There is a similar case for $j = n$.

Updates to the original sequence of $d \times d$ matrices can be translated directly into updates to the new sequence of $(d+1) \times (d+1)$ matrices. Queries to the original sequence result in at most four updates and a query on the new sequence. This completes the proof.
\end{proof}

\begin{restate}[Corollary~\ref{cor:matrix-prod-lower}]
  Theorem~\ref{thm:group_range_LB} holds for the Matrix Product Problem.
\end{restate}


\subsection{Upper Triangular Matrices} \label{subsec:upper-tri}

We further restrict our focus to the group $G$ of invertible upper triangular matrices. In some applications, only upper triangular matrices are sufficient instead of the full general linear group of all invertible matrices, and the proof of Lemma \ref{lem:mat-prod} does not immediately imply that the \emph{Upper Triangular Matrix Product Problem} has a Theorem \ref{thm:group_range_LB} style of lower bound, as our gadget would make one matrix no longer upper triangular. Nonetheless, we are able to prove the lower bound via a modification of Lemma \ref{lem:mat-prod}.

\begin{lemma} \label{lem:mat-prod-triang}
If the Upper Triangular Matrix Product Problem for $n$ matrices of dimension $d \times d$ can be solved in amortized $T(n,d)$ time per operation, then the Matrix Range Problem $n$ matrices of dimension $d \times d$ can be solved in amortized $O(T(2n+1,2d))$ time per operation.
\end{lemma}

\begin{proof}
The reduction uses the following identity: Let $N_{(i,j)}$ denote the $d \times d$ matrix which has all entries 0 except its $(i,j)$ entry is $1$. For any $d \times d$ upper triangular matrices $A$, $B$, and $C$, we have the following identity of $2d \times 2d$ upper triangular matrices:
\[
  \fourmat{A}{0}{0}{I}
  \fourmat{N_{(1,j)}}{0}{0}{I}
  \fourmat{B}{0}{0}{I}
  \fourmat{I}{N_{(i,1)}}{0}{I}
  \fourmat{C}{0}{0}{I}
	=
	\fourmat{AB_jC}{B_{(i,j)}}{0}{I},
\]
where $B_j$ is the all zeroes matrix except that its first row is the $j$th row of $B$, and $B_{(i,j)}$ is the all zeroes matrix except that its top right entry is the $(i,j)$ entry of $B$.

Similar to before, to maintain the sequence $M_1, \ldots, M_n$ of $d \times d$ matrices, we will maintain the following sequence of $2d \times 2d$ matrices:
\[
  \fourmat{I}{0}{0}{I}
  \fourmat{M_1}{0}{0}{I}
  \fourmat{I}{0}{0}{I}
  \fourmat{M_2}{0}{0}{I}
  \fourmat{I}{0}{0}{I}
	\cdots
  \fourmat{I}{0}{0}{I}
  \fourmat{M_n}{0}{0}{I}
  \fourmat{I}{0}{0}{I}.
\]
To query the $(i,j)$ entry of the product $M_a M_{a+1} \cdots M_b$, we change the $(2a-1)$th matrix to $\fourmat{N_{(1,j)}}{0}{0}{I}$, and change the $(2b+1)$th matrix to $\fourmat{I}{N_{(i,1)}}{0}{I}$, and then our desired value is the top right entry of the product of all the matrices.

\end{proof}

\begin{corollary}
  Theorem~\ref{thm:group_range_LB} holds for the Upper Triangular Matrix Product Problem.
\end{corollary}

\ifx\undefined\mainfile
\end{document}
\fi

\ifx\undefined\mainfile







\fi

\section{Applications of the Group Range Problem} \label{app:app}

\subsubsection*{Physics}

One example in the areas of optics and computer graphics is the propagation of electromagnetic waves through different media. The \emph{transfer-matrix method} from optics describes how to analyze the propagation of such waves by computing a product of \emph{characteristic matrices}, one corresponding to each medium. In optical modeling experiments, physicists sometimes need to quickly determine how making changes to one characteristic matrix alters the overall product, a computational task described by our problem. Some forms of ray tracing in computer graphics also use this transfer matrix analysis~\cite{TM1, TM2, TM3}.

The transfer matrix method is also used in some mechanical engineering problems, like in the design of aircraft panels. Due to the details of these mechanics problems, the matrices involved are typically upper-triangular matrices. \cite{LinDonaldson}

In many of these applications, one is only interested in being able to query the product of the entire sequence of matrices, rather than querying arbitrary subintervals of matrices. We show that, when $G$ is the group of invertible matrices, or the group of invertible upper triangular matrices, our lower bound still holds even if only an entry of the product of the entire sequence of matrices can be queried.

\subsubsection*{Dynamic Permanent for Banded Matrices}


Consider the following data structure problem. We want to keep track of the
permanent of an $n \times n$ matrix $M$ over some finite field $\mathbb{F}_p$.
To keep the problem tractable (because Permanent is NP-complete), we restrict
attention to the case where $M$ is a band matrix, i.e. $M_{i, j}$ is nonzero
only when $|i - j| \le k$ for some constant $k$. We want to support the
following two operations:
\begin{itemize}
  \item Update($i, j, \Delta$), which updates
    $M_{i, j} \leftarrow M_{i, j} + \Delta$ but only for $|i - j| \le 1$.
  \item Query(), which returns the permanent of $M$.
\end{itemize}

It turns out that this problem is reducible to the Matrix Range Problem.
Consider the $k = 1$ case, and treat the permanent as the sum of weights of
perfect matchings of a bipartite graph $G = ([n], [n], E)$, we define $P_i$ to
be the sum of weights of perfect matchings of the bipartite graph
$G_i = ([i], [i], E \cap ([i] \times [i]))$. Because of the banded property of
the matrix, there are only two vertices that vertex $i$ on the left hand side of
the graph can be matched to: vertex $i-1$ or vertex $i$ on the right hand side.
Furthermore, if left $i$ is matched to right $i-1$ then left $i-1$ must be
matched to right $i$. Hence:
\begin{align*}
  P_i &= P_{i-1} M_{i, i} + P_{i-2} M_{i-1, i} M_{i, i-1} \\
  \left[
    \begin{array} {c}
      P_i \\
      P_{i-1}
    \end{array}
  \right]
    &=
  \left[
    \begin{array} {cc}
      M_{i, i} & M_{i-1, i} M_{i, i-1} \\
      1 & 0
    \end{array}
  \right]
  \left[
    \begin{array} {c}
      P_{i-1} \\
      P_{i-2}
    \end{array}
  \right] \\
  \left[
    \begin{array} {c}
      P_n \\
      P_{n-1}
    \end{array}
  \right]
    &=
  \left[
    \begin{array} {cc}
      M_{n, n} & M_{n-1, n} M_{n, n-1} \\
      1 & 0
    \end{array}
  \right]
    \cdots
  \left[
    \begin{array} {cc}
      M_{2, 2} & M_{1, 2} M_{2, 1} \\
      1 & 0
    \end{array}
  \right]
  \left[
    \begin{array} {c}
      M_{1, 1} \\
      1
    \end{array}
  \right]
\end{align*}

Therefore it suffices to keep a Matrix Range which stores the $n-1$ matrices
$\left[
    \begin{array} {cc}
      M_{i, i} & M_{i-1, i} M_{i, i-1} \\
      1 & 0
    \end{array}
  \right]$
. Each update to the Dynamic Permanent data structure results in exactly one
update to the Matrix Range data structure, and a query to the Dynamic Permanent
data structure can be answered by querying for the top row (two entries) of the
product of the entire range.

Modulo the fact that these matrices may not be invertible (if
$M_{i-1, i} M_{i, i-1} = 0$), our results show that this approach to the problem
should cost $\Omega(\log n)$ time per operation. In particular, our lower bound
for the Matrix Product Problem showed that this problem is still hard even when
queries only request the entire range of matrices and not arbitrary subintervals.

\ifx\undefined\mainfile
\end{document}
\fi

\end{document}